\documentclass[
    reprint,
    aps,
    prx,
	amsmath,
	amssymb,
	superscriptaddress,
	longbibliography,
	english
]{revtex4-2}

\usepackage{float}
\usepackage{physics}
\usepackage{graphicx}
\usepackage[dvipsnames]{xcolor}
\usepackage[separate-uncertainty = true,multi-part-units=single]{siunitx}
\usepackage[hidelinks, colorlinks=True, linkcolor=BlueViolet, filecolor=BlueViolet, urlcolor=BlueViolet, citecolor=BlueViolet]{hyperref}
\usepackage[all]{hypcap}
\usepackage[title]{appendix}

\DeclareSIUnit\phinaught{\Phi_0}

\begin{document}

\title{Parametric multi-element coupling architecture for coherent and dissipative control of superconducting qubits}
\author{G.~B.~P.~Huber}
\email{gerhard.huber@wmi.badw.de}
\affiliation{Technical University of Munich, TUM School of Natural Sciences, Department of Physics, Garching 85748, Germany}
\affiliation{Walther-Mei{\ss}ner-Institut, Bayerische Akademie der Wissenschaften, Garching 85748, Germany}

\author{F.~A.~Roy}
\affiliation{Walther-Mei{\ss}ner-Institut, Bayerische Akademie der Wissenschaften, Garching 85748, Germany}
\affiliation{Theoretical Physics, Saarland University, Saarbr\"{u}cken 66123, Germany}

\author{L.~Koch}
\affiliation{Technical University of Munich, TUM School of Natural Sciences, Department of Physics, Garching 85748, Germany}
\affiliation{Walther-Mei{\ss}ner-Institut, Bayerische Akademie der Wissenschaften, Garching 85748, Germany}

\author{I.~Tsitsilin}
\affiliation{Technical University of Munich, TUM School of Natural Sciences, Department of Physics, Garching 85748, Germany}
\affiliation{Walther-Mei{\ss}ner-Institut, Bayerische Akademie der Wissenschaften, Garching 85748, Germany}

\author{J.~Schirk}
\affiliation{Technical University of Munich, TUM School of Natural Sciences, Department of Physics, Garching 85748, Germany}
\affiliation{Walther-Mei{\ss}ner-Institut, Bayerische Akademie der Wissenschaften, Garching 85748, Germany}

\author{N.~J.~Glaser}
\affiliation{Technical University of Munich, TUM School of Natural Sciences, Department of Physics, Garching 85748, Germany}
\affiliation{Walther-Mei{\ss}ner-Institut, Bayerische Akademie der Wissenschaften, Garching 85748, Germany}

\author{N.~Bruckmoser}
\affiliation{Technical University of Munich, TUM School of Natural Sciences, Department of Physics, Garching 85748, Germany}
\affiliation{Walther-Mei{\ss}ner-Institut, Bayerische Akademie der Wissenschaften, Garching 85748, Germany}

\author{C.~Schweizer}
\affiliation{Walther-Mei{\ss}ner-Institut, Bayerische Akademie der Wissenschaften, Garching 85748, Germany}
\affiliation{Fakult\"{a}t f\"{u}r Physik, Ludwig-Maximilians-Universit\"{a}t M\"{u}nchen, Schellingstra{\ss}e 4, M\"{u}nchen 80799 , Germany}

\author{J.~Romeiro}
\affiliation{Technical University of Munich, TUM School of Natural Sciences, Department of Physics, Garching 85748, Germany}
\affiliation{Walther-Mei{\ss}ner-Institut, Bayerische Akademie der Wissenschaften, Garching 85748, Germany}

\author{G.~Krylov}
\affiliation{Technical University of Munich, TUM School of Natural Sciences, Department of Physics, Garching 85748, Germany}
\affiliation{Walther-Mei{\ss}ner-Institut, Bayerische Akademie der Wissenschaften, Garching 85748, Germany}

\author{M.~Singh}
\affiliation{Technical University of Munich, TUM School of Natural Sciences, Department of Physics, Garching 85748, Germany}
\affiliation{Walther-Mei{\ss}ner-Institut, Bayerische Akademie der Wissenschaften, Garching 85748, Germany}

\author{F.~X.~Haslbeck}
\affiliation{Technical University of Munich, TUM School of Natural Sciences, Department of Physics, Garching 85748, Germany}
\affiliation{Walther-Mei{\ss}ner-Institut, Bayerische Akademie der Wissenschaften, Garching 85748, Germany}

\author{M.~Knudsen}
\affiliation{Walther-Mei{\ss}ner-Institut, Bayerische Akademie der Wissenschaften, Garching 85748, Germany}
\affiliation{Technical University of Munich, TUM School of Computation, Information and Technology, Department of Computer Science, Garching 85748, Germany}

\author{A.~Marx}
\affiliation{Technical University of Munich, TUM School of Natural Sciences, Department of Physics, Garching 85748, Germany}
\affiliation{Walther-Mei{\ss}ner-Institut, Bayerische Akademie der Wissenschaften, Garching 85748, Germany}

\author{F.~Pfeiffer}
\affiliation{Technical University of Munich, TUM School of Natural Sciences, Department of Physics, Garching 85748, Germany}
\affiliation{Walther-Mei{\ss}ner-Institut, Bayerische Akademie der Wissenschaften, Garching 85748, Germany}

\author{C.~Schneider}
\affiliation{Technical University of Munich, TUM School of Natural Sciences, Department of Physics, Garching 85748, Germany}
\affiliation{Walther-Mei{\ss}ner-Institut, Bayerische Akademie der Wissenschaften, Garching 85748, Germany}

\author{F.~Wallner}
\affiliation{Technical University of Munich, TUM School of Natural Sciences, Department of Physics, Garching 85748, Germany}
\affiliation{Walther-Mei{\ss}ner-Institut, Bayerische Akademie der Wissenschaften, Garching 85748, Germany}

\author{D.~Bunch}
\affiliation{Technical University of Munich, TUM School of Natural Sciences, Department of Physics, Garching 85748, Germany}
\affiliation{Walther-Mei{\ss}ner-Institut, Bayerische Akademie der Wissenschaften, Garching 85748, Germany}

\author{L.~Richard}
\affiliation{Technical University of Munich, TUM School of Natural Sciences, Department of Physics, Garching 85748, Germany}
\affiliation{Walther-Mei{\ss}ner-Institut, Bayerische Akademie der Wissenschaften, Garching 85748, Germany}

\author{L.~S\"{o}dergren}
\affiliation{Technical University of Munich, TUM School of Natural Sciences, Department of Physics, Garching 85748, Germany}
\affiliation{Walther-Mei{\ss}ner-Institut, Bayerische Akademie der Wissenschaften, Garching 85748, Germany}

\author{K.~Liegener}
\affiliation{Technical University of Munich, TUM School of Natural Sciences, Department of Physics, Garching 85748, Germany}
\affiliation{Walther-Mei{\ss}ner-Institut, Bayerische Akademie der Wissenschaften, Garching 85748, Germany}

\author{M.~Werninghaus}
\affiliation{Technical University of Munich, TUM School of Natural Sciences, Department of Physics, Garching 85748, Germany}
\affiliation{Walther-Mei{\ss}ner-Institut, Bayerische Akademie der Wissenschaften, Garching 85748, Germany}

\author{S.~Filipp}
\email{stefan.filipp@wmi.badw.de}
\affiliation{Technical University of Munich, TUM School of Natural Sciences, Department of Physics, Garching 85748, Germany}
\affiliation{Walther-Mei{\ss}ner-Institut, Bayerische Akademie der Wissenschaften, Garching 85748, Germany}
\affiliation{Munich Center for Quantum Science and Technology (MCQST), Schellingstra{\ss}e 4, M\"{u}nchen 80799 , Germany}

\date{\today}

\begin{abstract}
As systems for quantum computing keep growing in size and number of qubits, challenges in scaling the control capabilities are becoming increasingly relevant.
Efficient schemes to simultaneously mediate coherent interactions between multiple quantum systems and to reduce decoherence errors can minimize the control overhead in next-generation quantum processors.
Here, we present a superconducting qubit architecture based on tunable parametric interactions to perform two-qubit gates, reset, leakage recovery and to read out the qubits.
In this architecture, parametrically driven multi-element couplers selectively couple qubits to resonators and neighbouring qubits, according to the frequency of the drive.
We consider a system with two qubits and one readout resonator interacting via a single coupling circuit and experimentally demonstrate a controlled-Z gate with a fidelity of \SI{98.3(0.23)}{\%}, a reset operation that unconditionally prepares the qubit ground state with a fidelity of \SI{99.8(0.02)}{\%} and a leakage recovery operation with a \SI{98.5(0.3)}{\%} success probability.
Furthermore, we implement a parametric readout with a single-shot assignment fidelity of \SI{88(0.4)}{\%}.
These operations are all realized using a single tunable coupler and a statically decoupled resonator, demonstrating the experimental feasibility of the proposed architecture and its potential for reducing the system complexity in scalable quantum processors.
\end{abstract}
\maketitle
\section{Introduction}
Superconducting qubits are a leading platform for quantum computing and for quantum simulations and have been utilized in recent experiments exploring the boundaries of useful quantum applications~\cite{Arute2019,Kim2023,Krinner2022,Acharya2023}.
However, as devices scale in size and number of qubits their successful function relies on overcoming technical challenges in both experimental infrastructure and qubit control.
The cooling power of the cryogenic system limits the total number of signal lines to the device~\cite{Krinner2019} and the circuit footprint limits the density of qubits on the chip.
Therefore, scalable architectures should minimize the quantity and size of structures supporting each qubit, such as control lines, readout resonators and coupling elements.
Still, the basic quantum computing operations of readout, state preparation and single- and two-qubit gates need to be implemented with high fidelity to reduce the accumulation of errors.
In particular, leakage to non-computational states can spread throughout the device, significantly increasing the failure rate of quantum error correction protocols~\cite{Suchara2015,Bultink2020,Varbanov2020,McEwen2021,Battistel2021,Miao2023,Krinner2022,Acharya2023}.
Furthermore, the passive initialization of qubits in larger devices is adversely impacted by the presence of thermal excitations~\cite{Buffoni2022} and the increasing coherence times of qubits~\cite{Wang2022,Place2021,Koch2024_inprep}.
Finally, the coupling between qubit and resonator, needed for performing qubit readout, limits the qubit lifetime due to the Purcell effect~\cite{Houck2008}, and introduces dephasing noise due to residual excitations in the resonator~\cite{Schuster2005,Clerk2007,Sears2012,Yan2018}.

Several solutions have been proposed to address these challenges.
To reduce the heat load of control signals, cryogenic electronics~\cite{Joshi2022, Mehrpoo2019} and photonic links~\cite{Lecocq2021, Delaney2022, VanThiel2023} as well as architectures with reduced number of control lines~\cite{Pechal2021} have been tested at small scales.
Similarly, resonators and qubits with reduced footprints have been fabricated using lumped-element designs~\cite{Zotova2023, Hazard2023, Zhao2020, Mamin2021, Bruckmoser2024, Wang2022_2, McFadden2025} and techniques to readout multiple qubits using the same resonator have been demonstrated~\cite{Filipp2009, Touzard2019, Noh2023}.
In addition, numerous superconducting non-pairwise couplers have been theoretically proposed ~\cite{Mezzacapo2014, Schondorf2019, Melanson2019, Menke2021, Kang2024, Simakov2024, Glaser2023}, with a recent experimental implementation relying on inductive couplings using flux qubits~\cite{Menke2022} and on static couplings in combination with a qudit to engineer local higher-order interactions~\cite{Zhang2022}. 
Different architectures have also been proposed, where multiple qubit modes are connected to common coupling elements, either via a shared connection to ground~\cite{Noh2023, Brown2022}, or as modular interconnects~\cite{Zhou2023}.

Moreover, active reset protocols have been implemented that initialize a qubit-state efficiently, typically by coupling the qubit to a lossy circuit element, such as a \SI{50}{\ohm} environment. 
Different methods have been introduced for this purpose: microwave drives~\cite{Egger2018, Magnard2018, Teixeira2024, Geerlings2013}, parametric flux drives on the qubit~\cite{Zhou2021, Nie2024} or adiabatic flux sweeps of the qubit~\cite{McEwen2021}.
Each of these methods has also been adapted to recover leakage out of the computational subspace by either driving different transitions~\cite{Battistel2021, Lacroix2023} or swapping excitations before applying a reset~\cite{Miao2023, chen2024}.
Purcell filters are commonly used~\cite{Reed2010_Fast, Jeffrey2014, Bronn2015, Heinsoo2018, Sunada2022, Sunada2024} to protect qubits from relaxation due to the coupling to the readout resonator.
To also protect from resonator-induced dephasing alternative decoupling approaches have been demonstrated, either utilizing multi-mode qubits~\cite{Gambetta2011, Gyenis2021, Dassonneville2020, Dassonneville2023, Pfeiffer2024} or parametric couplings~\cite{Noh2023, Yao2017, Allman2014}.

Here, we propose and implement an architecture, where a single coupling element controls the coherent interaction between qubits, their coupling to the readout resonator and their controlled dissipation into a lossy environment.
In contrast to previous methods, parametric drives are used to selectively activate not only two-qubit gates, but also qubit reset, leakage recovery and parametric qubit readout on a single device, all using the same coupler and control line.
This added capability reduces the overhead in supporting structures, providing a more scalable superconducting qubit architecture, while retaining functionality.
Now, the resonators serve a dual role, both as meters to determine the state of the qubits and as auxiliary elements to discard unwanted excitations.
Furthermore, compared to other architecture, the use of parametrically activated couplings helps to protect the qubits from photon-induced dephasing caused by spurious excitations of the readout resonator.
\begin{figure}[t]
\includegraphics{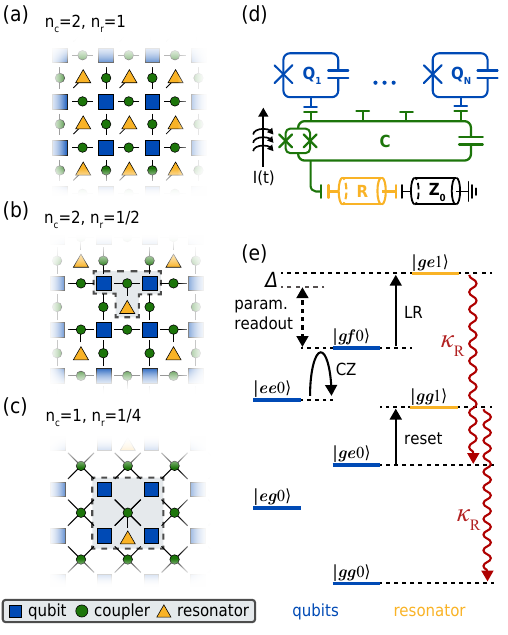}
\caption{\label{fig:1}
    Parametric coupling architecture.
    (a-c) Different architectures consisting of qubits Q (blue squares), flux-tunable couplers C (green circles) and resonators R (yellow triangles). The average number of couplers per qubit $n_c$ and resonators per qubit $n_r$ is given for each configuration.
    (a) Standard grid with static qubit-resonator interactions and coupler-mediated qubit-qubit interactions.
    (b) Parametric configuration with two-qubit couplers mediating interactions between qubit pairs and qubit resonators.
    (c) Four-qubit couplers create a high qubit connectivity and serve as mediators for the qubit resonator interaction.
    The grey box in (b) and (c) highlights the elemental unit of the parametric coupler architecture.
    (d) Simplified circuit of the elemental unit used for qubit-resonator interactions.
    Qubits (blue) and resonator (yellow) are capacitively connected to the coupler (green).
    The resonator is connected to the $Z_0 = \SI{50}{\ohm}$ waveguide impedance via an on-chip feedline. A time-dependent bias current $I(t)$ induces a flux $\varphi_\text{ext}(t)$ in the superconducting quantum interference device (SQUID) loop of the tunable coupler.
    (e) Energy levels of the qubit and resonator manifold with states labeled $\ket{\text{Q}_1\text{Q}_2\text{R}}$.
    The black arrows represent the transitions used for reset, leakage recovery, parametric readout and the controlled-Z gate.
    The resonator relaxation at a rate $\kappa_\text{R}$ is indicated by the red arrows.
    }
\end{figure}
\section{Parametric multi-element coupling architecture}
\label{sec:architecture}
In typical superconducting qubit architectures, qubits are statically coupled to individual readout resonators and interact with each other via static or tunable couplings.
This leads to a large overhead in the number of elements and control lines.
For instance, as shown in Fig.~\ref{fig:1}(a), a square lattice architecture with a tunable coupler (green circles) between each pair of qubits (blue squares) leads to $n_c =2$ additional coupling elements per qubit.
In this standard architecture there are $n_r=1$ readout resonators (yellow triangles) per qubit.
By extending the usage of the coupling elements to mediate not only the interaction between neighbouring qubits, but also between qubits and resonators, architectures with a severely reduced number of auxiliary elements can be envisaged, as shown in Fig.~\ref{fig:1}(b) and Fig.~\ref{fig:1}(c).
In the architecture in Fig.~\ref{fig:1}(b), each resonator is connected to two qubits, halving the required number of resonators to $n_r=1/2$.
This reduces the footprint per qubit and the number of readout lines.
The typical qubit measurement then requires a modification: the qubits are either read out in two sequential steps or pairs of qubits are measured simultaneously~\cite{Filipp2009,Touzard2019, Noh2023}.
The coupler functionality can be further adapted to couple more than two qubits, as depicted in Fig.~\ref{fig:1}(c) for a square-lattice with four qubits in each plaquette.
In this layout, each element is coupled to multiple qubits, further reducing the number of required couplers and resonators ($n_c = 1$, $n_r = 1/4$ for the square lattice).
In addition, the coupling of diagonally adjacent qubits in this configuration increases the qubit connectivity and opens up the possibility to implement multi-qubit gates using a single coupling element~\cite{Glaser2023, Menke2022}.

The elemental unit of these architectures, highlighted in grey boxes in Fig.~\ref{fig:1}(b,c) and sketched in circuit form in Fig.~\ref{fig:1}(d), consists of a flux-tunable coupler (C, green) capacitively connected to fixed-frequency transmon qubits (Q, blue) and a superconducting co-planar resonator (R, yellow).
The resonator couples via a microwave feedline to the environment, which has an impedance of $Z_0 = \SI{50}{\ohm}$.
The qubits and resonator couple strongly to the coupler and weakly to each other due to stray capacitances.
The Hamiltonian describing the system is given by
\begin{equation}
    \label{eq:Hamiltonian_system}
	\begin{split}
	\hat{H}/\hbar   &= \sum_{i\in{\text{\{Q,R,C\}}}} \omega_i \hat{a}_i^\dagger\hat{a}_i + \sum_{i \in \text{\{Q,C\}}} \frac{\alpha_i}{2} \hat{a}_i^\dagger\hat{a}_i^\dagger\hat{a}_i\hat{a}_i\\
	&+	\sum_{\substack{i,j \in\text{\{Q,R,C\}} \\i \neq j}} \frac{g_{ij}}{2}(\hat{a}_i^\dagger-\hat{a}_i)(\hat{a}_j^\dagger-\hat{a}_j),
    \end{split}
\end{equation}
where $\omega_i$ denotes the frequency of qubits, resonator and coupler, $\alpha_i$ the anharmonicity of qubits and coupler and $g_{ij}$ the static capacitive couplings between elements.
During the parametric drive a time-dependent current $I(t)$, shown in Fig.~\ref{fig:1}(d), is applied to the flux line of the coupler.
This current harmonically modulates the external flux through the SQUID loop of the coupler ${\varphi_\text{ext}(t) =\varphi_\text{dc} + a_\text{D} \sin(\omega_\text{D}t)}$, where $\varphi_\text{dc}$ denotes the static flux offset, $a_\text{D}$ the drive amplitude and $\omega_\text{D}$ the drive frequency.
This in turn modulates the coupler frequency ${\omega_\text{C} = \sqrt{8 E_J E_C}-E_C}$, since the Josephson energy of the coupler depends on the external flux as ${E_J = E_\Sigma \abs{\cos\qty(\varphi_\text{ext})}\sqrt{1+d^2 \tan^2 \qty(\varphi_\text{ext})}}$.
Here, $E_\Sigma$ is the sum of the Josephson energies of each junction in the SQUID loop, $d$ their asymmetry and $E_C$ denotes the coupler charging energy~\cite{Koch2007}.
Since the coupler frequency $\omega_\text{C}$ is not linear in external flux, we expand it using a general Fourier series~\cite{Didier2018},
\begin{equation}
\omega_\text{C}(t) = \overline{\omega}_\text{C}+\sum_{m=1} D_m\cos\qty[m\qty(\omega_\text{D}t-\frac{\pi}{2})],
\end{equation}
where $\overline{\omega}_\text{C}$ denotes the average coupler frequency during the drive and the Fourier coefficients are given to first order by ${D_m \propto \frac{\partial^m \omega_\text{C}}{{\partial \varphi}^m}\big|_{\varphi_\text{dc}}} a_\text{D}^m$.
When the $k$-th harmonic of the drive frequency $k \omega_\text{D}$ is resonant with a transition between either two qubits or a qubit and the resonator, the drive activates a coupling with the strength
\begin{equation}
    \label{eq:g_param}
        \abs{\tilde{g}_{ij}} \approx \frac{g_{i\text{C}}g_{j\text{C}}}{k\omega_\text{D}}J_1(D_k/k\omega_\text{D}) \propto a_\text{D}^k,
\end{equation}
where $J_1$ is a Bessel function of the first kind.
Here, we use a Floquet transformation~\cite{Schweizer2019} to work in the frame of the drive and apply a Jacobi-Anger expansion.
Assuming small drive amplitudes, such that $D_m\ll m\omega_\text{D}$ for all $m\le k$, we find the coupling strength to be linearly proportional to the $k$-th Fourier coefficient $D_k$ of the drive, which is proportional to the $k$-th power of the drive amplitude, $a_\text{D}^k$.
See Appendix~\ref{app:eff_ham} for the full derivation.

By tuning the parametric drive frequency and its amplitude we can, therefore, selectively activate and control couplings between specific states of the system [Fig.~\ref{fig:1}(e); black arrows].
We label the states of the qubit as $\ket{g}$, $\ket{e}$ and $\ket{f}$ for the ground, first and second-excited states, respectively, and the states of the resonator as $\ket{n}$, where $n$ denotes the photon number.
The {controlled-Z gate} is performed by driving the $\ket{ee} \leftrightarrow \ket{fg}$ transition between two qubits~\cite{Caldwell2018, Reagor2018, Ganzhorn2020, Hong2020, Sete2021parametric}.
To realize a qubit reset, we drive the transition between $\ket{e0}$ and $\ket{g1}$ to transfer the excitation in the qubit to the resonator, where it rapidly decays into the \SI{50}{\ohm} environment.
Similarly, driving between $\ket{f0}$ and $\ket{e1}$ implements the leakage recovery operation.
By parametrically driving the same transition with a small detuning $\Delta$ a qubit state dependent shift of the resonator is generated, which is used for the parametric readout.

To experimentally demonstrate the four operations of reset, leakage recovery, parametric readout and two-qubit gate, we use a superconducting qubit device with three fixed-frequency qubits ($\text{Q}_1, \text{Q}_2, \text{Q}_3$) and a co-planar resonator (R) all capacitively coupled to the same flux-tunable coupling element (C).
We perform the two-qubit gate on the pair $\text{Q}_1\text{Q}_2$ and the other operations on $\text{Q}_1$.
For the characterization of these operations we utilize standard dispersive readout on auxiliary resonators.
More details on the sample, the experimental setup and control pulses are provided in Appendix~\ref{app:exp_setup}.
\section{Qubit reset}
\begin{figure}[t]
\includegraphics{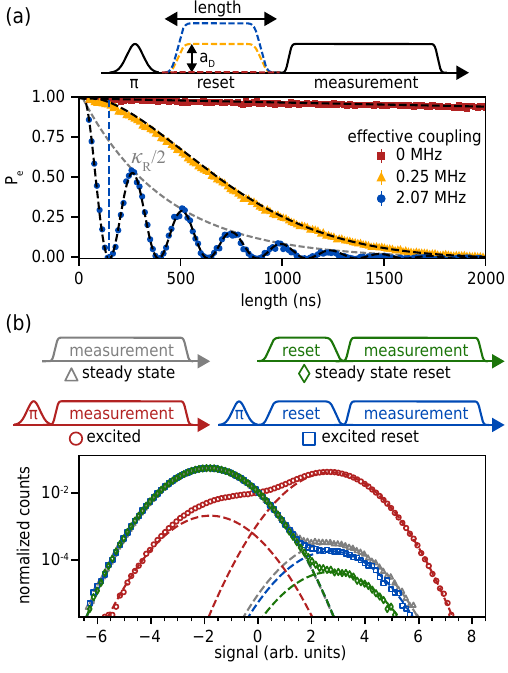}
\caption{\label{fig:reset}
    (a) Excited state population $P_e$ of the qubit during the flux drive for zero parametric coupling (red squares), for weak coupling (yellow triangles) and for strong coupling (blue circles).
    The grey dashed line shows the decay envelope $\kappa_\text{R}/2$ for the underdamped system.
    The length of the reset pulse is marked by the vertical dashed line at \SI{150}{\nano\second}.
    Simulated dynamics are shown as black dashed lines.
    (b) Pulse schemes and histogram of the integrated measurement signal for different experiments: measurement only (grey triangles), reset pulse (green diamonds), $\pi$-pulse (red circles), $\pi$-pulse followed by a reset pulse (blue squares).
    The width and position of the two Gaussian distributions for ground and excited state preparation (dashed lines) are fitted once and kept constant for the four experiments, with the distribution height as the only free fit parameter.
}
\end{figure}
We perform a qubit reset by swapping unwanted excitations from the qubit ($\omega_{\text{Q}_1}/2\pi= \SI{3.83}{\giga\hertz}$), which has an intrinsic decay rate of $\Gamma_1/2\pi= \SI{4.2(1.6)}{\kHz}$, to the short-lived resonator ($\omega_{\text{R}}/2\pi= \SI{5.85}{\giga\hertz}$), which decays at the higher rate $\kappa_\text{R}/2\pi = \SI{770(3)}{\kHz}$.
We drive the coupler at the second-harmonic ($k=2$) of the $\ket{e0}\leftrightarrow\ket{g1}$ transition with a frequency of $\omega_{\text{D}}/2\pi= \SI{1.014}{\giga\hertz}$ which activates an effective coupling between qubit and resonator.
To calibrate the reset parameters we prepare the excited state of the qubit and measure its population as a function of time for different drive amplitudes (see Appendix~\ref{app:flux_pulse_calibration} for calibration) and fit the effective coupling, as shown in Fig.~\ref{fig:reset}(a).
Without the reset pulse the qubit decays at its intrinsic relaxation rate $\Gamma_1$ (red squares).
Increasing the amplitude of the applied parametric drive changes the system dynamics from overdamped (yellow triangles) to underdamped (blue circles), where the system relaxes at the higher rate $\sim\kappa_\text{R}/2$ (derivation in Appendix~\ref{app:dynamic_decay}).
Here, two types of reset can be realized:
Either, the excitation is swapped from the qubit to the resonator, where it decays at the intrinsic resonator decay rate $\kappa_\text{R}$, or the system is continuously driven until both qubit and resonator have relaxed with the joint decay rate $\kappa_\text{R}/2$.
While the latter is robust against small fluctuations in pulse parameters, it comes at the cost of an increased pulse length and overall reset length.
Therefore, we here choose to use the former method to implement the reset operation.
By driving with an effective qubit resonator coupling of \SI{2.07}{\MHz} the qubit excitation fully swaps to the resonator within $\tau_{\text{r}}=\SI{150}{\nano\second}$.
Once swapped, the resonator relaxes at the full decay rate $\kappa_\text{R}$, completing the reset.

Next, we determine the effect of the calibrated reset pulse on the state of the qubit by comparing single-shot measurement outcomes for different preparation sequences [Fig.~\ref{fig:reset}(b)].
Each of the four measurements consists of $26\times2^{17}$ shots, with a delay between individual shots of $\SI{500}{\micro\second}\gg 1/\Gamma_1$ to ensure independent results from consecutive experimental realizations.

To determine the reset efficiency $\eta_\text{r}=1-P_{\pi}^\text{r}/P_{\pi}$ we compare the excited state population after a $\pi$-pulse with the reset applied ${P_{\pi}^\text{r}=\SI{0.33(0.03)}{\%}}$ to the excited state population without reset ${P_{\pi}=\SI{88(0.3)}{\%}}$, shown in Fig.~\ref{fig:reset}(b) as blue squares and red circles, respectively.
The measured efficiency $\eta_\text{r}=\SI{99.63(0.03)}{\%}$ sets the upper bound for the excited state population after a single reset operation $P^\text{r}\leq1-\eta_\text{r}$.
To determine the success probability in preparing the ground state from any initial state, we compute the reset fidelity $\mathcal{F}_\text{r}=1-\qty(P_{\text{Id}}^\text{r}+P_\pi^\text{r})/2=\SI{99.8(0.02)}{\%}$, using the residual excited population when no pulses are applied $P_{\text{Id}} = \SI{0.62(0.04)}{\%}$ [Fig.~\ref{fig:reset}(b); grey triangles] and when the reset is applied $P_{\text{Id}}^\text{r}=\SI{0.074(0.005)}{\%}$ [Fig.~\ref{fig:reset}(b); green diamonds].

Assuming a thermal distribution, the reset corresponds to a cooling of the qubit from its initial temperature $\text{T}_{\text{Id}}=\SI{36.3(0.4)}{\milli \kelvin}$ to a temperature after reset of $\text{T}^\text{r}_{\text{Id}}=\SI{25.5(0.2)}{\milli \kelvin}$.
Given the achieved reset efficiency one would expect an even stronger cooling power.
However, the qubit population after the reset is bound from below by the thermal population in the resonator $ n_\text{th} $ and the non-zero population $n_\uparrow$ in the qubit due to thermalization during the reset and measurement pulses.
Assuming the resonator couples to the same thermal environment as the qubit yields an average resonator photon number of $n_\text{th}= \SI{0.047(0.004)}{\%}$.
In general, even for a perfect swap the temperature of the qubit after reset is bounded from below as $\text{T}^\text{r}_{\text{Id}} \ge \omega_q/\omega_r \text{T}_\text{R}$, where $\text{T}_\text{R}$ denotes the resonator temperature~\cite{Grajcar2008}.
Additionally, the qubit thermalizes at a rate $\kappa_{0\rightarrow 1} \approx P_{\text{Id}}\Gamma_1$ yielding $n_\uparrow \approx \kappa_{0\rightarrow 1} \frac{\tau_{\text{r}}+\tau_{\text{m}}}{2} = \SI{0.020(0.008)}{\%}$, where $\tau_{\text{r}} = \SI{150}{\nano\second}$ denotes the reset length and $\tau_{\text{m}} = \SI{2.3}{\micro\second}$ the measurement length~\cite{Magnard2018}.
The sum of these two populations gives a lower bound on the measured qubit population after the applied reset pulse $n_\text{th} + n_\uparrow =\SI{0.067(0.009)}{\%}$.
This limit agrees within the error bounds with the experimentally observed population $P_{\text{Id}}^\text{r}$, showing that the observed cooling power of the reset is limited by the relatively low frequency of the resonator and thermalization of the qubit during measurement.
\section{Leakage recovery}
\begin{figure}[th!]
\includegraphics{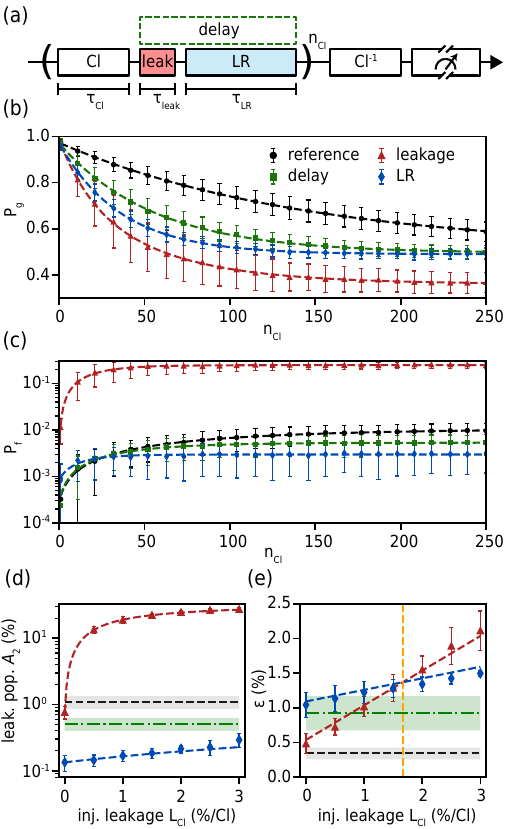}
\caption{\label{fig:LR}
    (a) Pulse sequence for the leakage randomized benchmarking (RB) experiment containing Clifford gates (Cl), induced leakage (leak), and the leakage recovery operation (LR).
    (b) Population $P_g$ in the ground state $\ket{g}$ as a function of the number of Clifford gates $\text{n}_\text{Cl}$ for different interleaved randomized benchmarking protocols: reference (black circles), \SI{2}{\%} leakage operation (red triangles), \SI{2}{\%} leakage operation followed by a LR operation (blue diamonds) and a delay with the same length as leakage and LR operation combined (green squares). Dashed lines are leakage-RB fits.
    (c) Leakage population in the $\ket{f}$-state extracted from similar measurements as described in (b). 
    (d) Measured equilibrium leakage population $A_2$ and (e) average gate error $\varepsilon$ as a function of the injected leakage $L_\text{Cl}$.
    Dashed lines are predictions from rate equations and the error rate, respectively (see main text for details). The vertical yellow line in (e) shows the break-even point between the error from leakage and from decoherence during the LR operation.
}
\end{figure}

Since transmon qubits are not true two-level systems, leakage out of the computational subspace can occur during device operation~\cite{Suchara2015,Bultink2020, Varbanov2020}, even when using mitigation techniques such as pulse-shaping~\cite{Motzoi2009,Rol2019}.
To recover excitations leaked to higher qubit states we adapt the technique already used in the reset protocol.
Instead of the $\ket{e}$-state, we now couple the resonator to the second-excited $\ket{f}$-state of the transmon to introduce a fast relaxation channel for the leaked excitations.
This dissipative decay mechanism removes the phase information of the excited state and transforms a leakage error into a Pauli error within the computational subspace, which is less detrimental for computational tasks, particularly in the context of quantum error correction~\cite{Fowler2013, Suchara2015, Bultink2020}.
We engineer the required relaxation channel by driving the second-harmonic ($k=2$) of the $\ket{f0} \leftrightarrow \ket{e1}$ transition ($\omega_\text{D}/2\pi = \SI{1.112}{\GHz}$).
For an effective coupling of \SI{0.91}{\MHz} the population in $\ket{f0}$ is fully transferred to $\ket{e1}$ within $\tau_\text{LR}=\SI{310}{\nano\second}$.
This state then decays to $\ket{e0}$ at the resonator decay rate $\kappa_\text{R}$, completing the leakage recovery (LR).
We apply the LR to the qubit prepared in the second-excited state and measure a residual leakage population of $P_f^\text{LR} = \SI{1.5(0.3)}{\%}$, yielding a recovery success probability of $\mathcal{F}_\text{LR} = 1-P_f^\text{LR}= \SI{98.5(0.3)}{\%}$.
The parametric drive shifts the effective frequency of the qubit during the protocol, leading to an unwanted additional phase on the qubit.
We correct for this additional phase by applying a virtual-Z gate~\cite{McKay2017} after the parametric drive, ensuring that the LR operation produces minimal impact on the computational basis states, as shown in Appendix~\ref{app:LR_cal}.\\
To evaluate the performance of the LR operation we perform a randomized benchmarking (RB) protocol, consisting of sequences of random Clifford gates~\cite{Knill2008, Magesan2011, Wood2018}.
Between each Clifford gate we inject a controlled amount of leakage $L_\text{Cl}$ by applying a weak pulse resonant with the ${\ket{e}\leftrightarrow\ket{f}}$ transition, followed by a LR operation, as shown in the pulse sequence in Fig.~\ref{fig:LR}(a).
We measure the ground state population $P_g$ [Fig.~\ref{fig:LR}(b)] and the second-excited state population $P_f$ [Fig.~\ref{fig:LR}(c)] of the qubit as a function of the number of Clifford gates $\text{n}_\text{Cl}$ (see Appendix~\ref{app:gef_readout} for details on readout).
We apply exponential fits to the data $P_g = A_0 + B_0\lambda_0^{\text{n}_\text{Cl}} + B_2 \lambda_2^{\text{n}_\text{Cl}}$ and $P_f = A_2 + B_2 \lambda_2^{\text{n}_\text{Cl}}$, where $A_2$ [Fig.~\ref{fig:LR}(d)] specifically captures the equilibrium leakage population reached in the steady-state~\cite{Wood2018,Werninghaus2021}.
From these we extract the average gate error $\varepsilon = (1-\lambda_0+L_2)/2$ [Fig.~\ref{fig:LR}(e)], where $L_2 = (1-A_2)(1-\lambda_2)$ is the leakage rate given by the injected leakage $L_\text{Cl}$.
We perform first a reference RB measurement based on calibrated single-qubit gates [Fig.~\ref{fig:LR}(b,c); black circles] and obtain an equilibrium leakage of $A_2^\text{ref}=\SI{1.1(0.2)}{\%}$ and an error of $\varepsilon_\text{ref}=\SI{0.34(0.09)}{\%}$ [Fig.~\ref{fig:LR}(d,e); black dashed line].
The measured error agrees within error bounds to the prediction due to decoherence $\Gamma_\Sigma\tau_\text{Cl}/3 = \SI{0.36(0.09)}{\%}$~\cite{Wood2018}, where $\Gamma_\Sigma$ is the sum of the decay rate $\Gamma_1$ and the pure dephasing rate $\Gamma_\phi$ and $\tau_\text{Cl}=\SI{200}{\ns}$ is the average length of one Clifford gate.
Injecting a fixed amount of leakage $L_\text{Cl}=\SI{2}{\%}$ increases the population in $\ket{f}$, causing the population in the ground state $\ket{g}$ to converge to a value $< 0.5$ [Fig.~\ref{fig:LR}(b,c); red triangles].
The equilibrium leakage $A_2^\text{leak}$ [Fig.~\ref{fig:LR}(d); red triangles] then depends on the injected leakage $L_\text{Cl}$ according to the equation
\begin{equation} 
\label{eq:A_2_no_LR}
A_2^\text{leak} = \frac{L_\text{Cl}}{3L_\text{Cl}+2\qty[e^{\tau_\text{Cl}\Gamma_{f\rightarrow e}}-1]},
\end{equation}
which accounts for relaxation of the $\ket{f}$-state to $\ket{e}$-state at the decay rate $\Gamma_{f\rightarrow e} = \SI{6.3(0.7)}{\kHz}$ (Appendix~\ref{app:rates}).
As expected from a qutrit leakage model, in the limit of no qubit decay the population in the second-excited state $A_2^\text{leak}$ saturates at $1/3$.
The injected leakage introduces an additional leakage error, increasing linearly with $L_\text{Cl}$ as~\cite{ Abad2022}
\begin{equation} 
\label{eq:error_no_LR}
\varepsilon_\text{leak} = \frac{\Gamma_\Sigma(\tau_\text{Cl}+\tau_\text{leak})}{3} + \frac{L_\text{Cl}}{2},
\end{equation}
where $\tau_\text{leak}=\SI{100}{\ns}$ is the length of the leakage injection.
The introduction of the LR operation recovers the leaked qubit population and therefore suppresses leakage accumulation, successfully limiting the equilibrium leakage $A_2^\text{LR}$ below \SI{0.3}{\%} for all tested injected leakage values [Fig.~\ref{fig:LR}(b-d); (blue diamonds].
From the assumption that the LR operation reduces the leakage rate by $(1-\mathcal{F}_\text{LR})$ one would expect an even lower equilibrium leakage.
However, the finite decay rate of the resonator can not be neglected, as the resonator does not fully deplete between repeated LR operations.
Taking this additional effect into account leads to a modified equation
\begin{equation}
\label{eq:A_2_LR_Assuming_TR_<Tef}
A_2^\text{LR} = \frac{L_\text{Cl} e^{-(\tau_\text{Cl}+\tau_\text{leak}+\tau_\text{LR})\kappa_\text{R}} }{3 L_\text{Cl} + 4\qty[1- e^{-(\tau_\text{Cl}+\tau_\text{leak}+\tau_\text{LR})\kappa_\text{R}}]},
\end{equation}
where we assume a perfect LR operation $\mathcal{F}_\text{LR}=1$ and no second-excited state decay $\Gamma_{f\rightarrow e} = 0$ (derivation in Appendix~\ref{app:rates}), agreeing with the measurement.
Since the LR operation relies on a dissipative process, leakage errors are transformed into Pauli errors in the form of dephasing within the qubit subspace.
This reduces the scaling of the overall error as a function of $L_\text{Cl}$ [Fig.~\ref{fig:LR}(e); blue diamonds] by a factor of three compared to Eq.~\eqref{eq:error_no_LR}
\begin{equation} 
\label{eq:error_LR}
\varepsilon_\text{LR} = \frac{\Gamma_\Sigma(\tau_\text{Cl}+\tau_\text{leak}+\tau_\text{LR})}{3} + \frac{L_\text{Cl}}{6},
\end{equation}
as derived in Appendix~\ref{app:rates}.
This improvement in scaling is offset by the decoherence occurring during the LR operation itself $\Gamma_\Sigma\tau_\text{LR}/3$.
Within uncertainty bounds the same increase can also be observed in $\varepsilon$ when the LR operation is replaced with a delay of equal length [Fig.~\ref{fig:LR}(e); green dashed-dotted line], suggesting that the error from the LR operation is decoherence limited.
Therefore, while the application of the LR operation is always advantageous to suppress the accumulation of leakage, it also manages to reduce the overall error when $\varepsilon_\text{LR} < \varepsilon_\text{leak}$ which occurs when $\tau_{\text{LR}}\Gamma_\Sigma < L_\text{Cl}$ [Fig.~\ref{fig:LR}(e); vertical yellow dashed line].
\section{Parametric Readout}
\begin{figure}[H]
\includegraphics{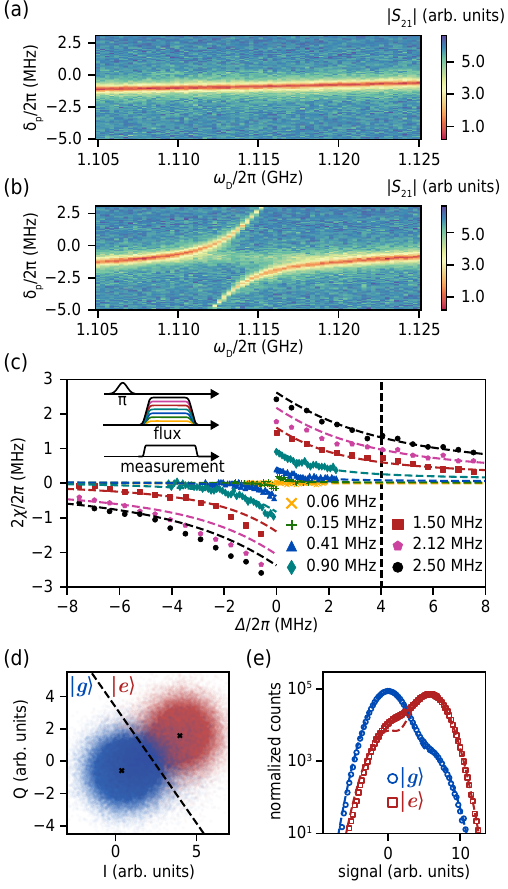} 
\caption{\label{fig:readout}
    Measured amplitude of the transmitted signal $|S_{21}|$ of the resonator as a function of parametric drive frequency $\omega_\text{D}$ and detuning $\delta_p = \omega_p-\omega_\text{R}$ of the probe tone from the undriven resonance frequency $\omega_\text{R}$, with the qubit being prepared either (a) in its ground state $\ket{g}$ or (b) in its excited state $\ket{e}$, for an effective coupling of $\tilde{g}_\text{QR}/2\pi =\SI{2.12(0.06)}{\MHz}$. 
    (c) Difference between the resonator frequency for the qubit being in $\ket{g}$ and $\ket{e}$ for varying parametric coupling strength $\tilde{g}_\text{QR}/2\pi = \{0.06, 0.15, 0.41, 0.90, 1.50, 2.12, 2.50\}$~\SI{}{\MHz} and the detuning $\Delta$ of the parametric drive from the resonance frequency in the frame of the drive.
    The vertical dashed line indicates the chosen optimal detuning for the parametric readout.
    The dashed lines are fits to the avoided crossings, see main text for details.
    (d) Single-shot data of the measured parametric readout of ground $\ket{g}$- and excited $\ket{e}$-state in the in-phase-quadrature (IQ) plane at a detuning $\Delta/2\pi = \SI{4.02}{\mega\hertz}$ and an effective coupling of $\tilde{g}_\text{QR}/2\pi =\SI{2.12(0.06)}{\MHz}$.
    (e) Projected histogram of parametric measurement. The dashed lines are Gaussian fits to the signal distributions with the qubit prepared in $\ket{g}$ and $\ket{e}$.
}
\end{figure}

During idling and gate operation the qubit is decoupled from the resonator, protecting it from dissipation into the environment and photon-induced dephasing.
Applying the parametric drive then activates the coupling and leads to a qubit-state dependent shift of the resonator frequency, which we utilize to read out the state of the qubit~\cite{Blais2004, Noh2023}.
We probe the resonator transmission near resonance, $\omega_p = \omega_\text{R} + \delta_p$, while simultaneously driving the $\ket{e, 1}\leftrightarrow\ket{f, 0}$ transition, as shown in Fig.~\ref{fig:readout}(a,b).
The slight increase of the resonance frequency in the measured range is a side effect of an avoided crossing between the tunable coupler and the resonator occurring at the drive frequency $\sim\SI{1.08}{\GHz}$, outside of the shown range.
Since the transition including the qubit ground state is off-resonant, the resonator and qubit hybridize if the qubit is in the excited state and not, if the qubit is in the ground state.
Therefore, we obtain a qubit-state dependent shift in the resonator frequency during the parametric drive, given by
\begin{equation}
\label{eq:avoided_crossing_chi}
   2\chi/2\pi = \frac{\qty|\Delta|-\sqrt{4 \qty|\tilde{g}_\text{QR}|^2 + \Delta^2}}{2},
\end{equation}
corresponding to an avoided crossing analogous to a drive-induced vacuum Rabi splitting~\cite{Haroche1989}.
Here, $\tilde{g}_\text{QR}$ denotes the effective qubit-resonator coupling from Eq.~\eqref{eq:g_param}, $\Delta = k \omega_\text{D}-(\tilde{\omega}_\text{R}-\tilde{\omega}_\text{Q}+\tilde{\alpha}_\text{Q})$ is the detuning of the $k$-th drive harmonic from the $\ket{f, 0}\leftrightarrow\ket{e, 1}$ transition and $\tilde{\omega}_\text{R}$, $\tilde{\omega}_\text{Q}$ and $\tilde{\alpha}_\text{Q}$ take into account drive-induced frequency shifts.
Note that, since the second-harmonic of the flux modulation is used ($k=2$), the detuning in the frame of the drive is twice the detuning in the lab frame.
In contrast to a standard readout scheme relying on a static capacitive coupling of qubit and resonator, both the magnitude and sign of the $\chi$-shift can be controlled by the frequency and amplitude of the parametric drive, which set the detuning $\Delta$ and the effective coupling $\tilde{g}_\text{QR}$ [Fig.~\ref{fig:readout}(c)].
We obtain a maximum frequency shift of $\abs{2\chi}/2\pi =\SI{2.5(0.12)}{\mega\hertz}$ at the avoided crossing ($\Delta = 0$).
From a measurement with the drive off we find a static contribution $2\chi_
\text{static}/2\pi=\SI{-7.6(0.4)}{\kilo\hertz}$, which is negligible when compared to the parametrically-induced $\chi$-shift.
We determine the assignment fidelity for varying drive parameters and find an optimum at a parametric coupling strength of $\tilde{g}_\text{QR}/2\pi = \SI{2.12(0.06)}{\MHz}$ and a detuning of $\Delta/2\pi=\SI{4.02}{\mega\hertz}$.
These parameters result in a shift of $\chi/2\pi=\SI{-0.53}{\mega\hertz}$, close to the {theoretical} optimum of $\abs{\chi}=\kappa_\text{R}/2$~\cite{Gambetta2008}. 
As $\tilde{g}_\text{QR}/\Delta \approx 0.53$ the dispersive approximation is no longer strictly valid and a part of the population in $\ket{e}$ is transferred to $\ket{f}$.
A subsequent application of the LR operation after the parametric readout and depletion of the resonator can bring the $\ket{f}$-state population back to the computational subspace.
We apply the measurement signal for $\tau_\text{meas}=\SI{10}{\micro\second}$ and obtain distinct measurement outcomes for the two qubit states, as can be observed in Fig.~\ref{fig:readout}(d) from the distribution of the in-phase and quadrature (IQ) components of the corresponding signals.
The resulting assignment fidelity is $\mathcal{F}_\text{meas} = \qty[P(g|g)+P(e|e)]/2 =\SI{88(0.4)}{\%}$, where $P(g|g)$ and $P(e|e)$ are the probabilities to correctly identify the ground state and excited state, respectively.

To understand the limit of the parametric readout fidelity, we fit Gaussian distributions to the IQ data of the ground and excited states and find an overlap assignment fidelity of $\mathcal{F}_\text{overlap} = \SI{95.0}{\%}$.
Additionally, the qubit decay from $\ket{e}$ to $\ket{g}$ during the measurement degrades the assignment fidelity by skewing the readout signal of the excited qubit state.
Assuming that the resonator reaches equilibrium instantaneously, an error occurs if the initially excited qubit is in the ground state for more than half the duration of the measurement.
Therefore, we estimate the probability $P(e|e)$ due to decay as $e^{- \tau_\text{meas} \Gamma_1 /2}$ leading to a fidelity of $\mathcal{F}_\text{decay}= \SI{93.9(2.3)}{\%}$.
Together, the error due to qubit decay during the measurement and the error due to finite signal overlap account for an expected parametric readout fidelity of $\mathcal{F}_\text{overlap}\cdot\mathcal{F}_\text{decay}=\SI{89.2(2.2)}{\%}$ matching with the measured value of $\mathcal{F}_\text{meas}$.
While in principle the measurement fidelity should be enhanced by even larger effective couplings, we find that for larger parametric drive amplitudes the effective resonator frequency shift is reduced and the resonator decay rate $\kappa_\text{R}$ is increased, which leads to a lower readout fidelity.
Similar to higher-order effects observed in other experiments utilizing parametric drives~\cite{Roth2017}.
While the impact of strong parametric drives has been analytically studied~\cite{Petrescu2023, Lagemann2022, Xiao2022}, a detailed investigation of the limits of the parametric readout is subject to forthcoming experiments.

\begin{figure}[t!]
\includegraphics{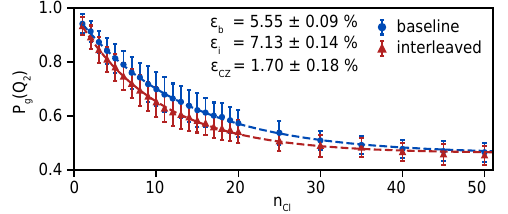} 
\caption{\label{fig:CZ}
Two-qubit randomized benchmarking without (blue circles) and with (red triangles) an interleaved controlled-Z gate.
The ground state population $P_g$ of qubit 2 is measured for 100 Clifford sequence randomizations each of length $\text{n}_\text{Cl}$. 
The average error of the controlled-Z gate $\varepsilon_\text{CZ}$ is extracted from the error per Clifford of the baseline RB $\varepsilon_\text{b}$ and the error per Clifford of the interleaved RB $\varepsilon_\text{i}$. 
}
\end{figure}
\section{Two-Qubit Gate}
To demonstrate the capability of the parametric coupling architecture to also realize two-qubit gates, we implement a controlled-Z gate between qubits $\text{Q}_1$ and $\text{Q}_2$ ($\ket{\text{Q}_1\text{Q}_2}$) following~\cite{Ganzhorn2020, Reagor2018, Hong2020, Sete2021parametric}.
We first excite both qubits and drive with a parametric pulse around the resonant frequency of the $\ket{ee} \leftrightarrow \ket{fg}$ transition.
For varying drive frequencies we use Ramsey sequences to determine the phase acquired after a full oscillation by the states $\ket{ee}$ and $\ket{ge}$.
We obtain a phase difference of $\pi$ with a drive duration of $\tau_\text{CZ} = \SI{339}{\nano\second}$ and a drive frequency of $\omega_\text{D} = \SI{489.89}{MHz}$ (see Appendix~\ref{app:CZ_calib}).
We calibrate and apply an additional virtual-Z gate~\cite{McKay2017} on both qubits after the parametric pulse to account for the qubit frequency shift during the parametric drive.
To determine the fidelity of the controlled-Z gate we perform a two-qubit interleaved randomized benchmarking experiment~\cite{Magesan2011}.
From a fit to the resulting curves shown in Fig.~\ref{fig:CZ} we extract a controlled-Z gate fidelity of $\mathcal{F}_{\text{CZ}} = 1 - \varepsilon_{\text{CZ}} = \SI{98.3(0.23)}{\%}$.
The gate fidelity agrees within error with the coherence limit $\mathcal{F}_{\text{CZ}}^{\text{limit}} = \SI{97.9(0.7)}{\%}$ imposed by the relaxation and dephasing rates of the two qubits~\cite{Abad2022}, implying that decoherence is the main error source.
\section{Discussion and Outlook}
\label{sec:discussion}
In the presented superconducting qubit architecture we utilize couplers that mediate interactions not only between two neighbouring qubits, but also between a qubit and a resonator.
This, in combination with the dual role played by a shared resonator, acting both as a dump for unwanted qubit excitations and as a readout element, reduces the overall number of resonators and couplers per qubit.
It also lowers the total number of control lines and the overall device footprint, while still allowing for all basic operations of a quantum processor.

In this architecture we perform a \SI{150}{\nano\second}-long reset operation preparing the qubit ground state with a fidelity of $\mathcal{F}_\text{r}=\SI{99.8(0.02)}{\%}$. 
The demonstrated protocol can be extended to also deplete higher excited levels by either introducing additional drive tones~\cite{Magnard2018, Egger2018, Zhou2021}, or by applying the LR operation and reset subsequently.
The reset operation fidelity can be increased further by raising the resonator frequency, which would reduce the residual excitation in the resonator, together with improved qubit coherence times, which would lead to less thermalization during reset.
Moreover, we implement a leakage recovery operation that reduces leakage by $\mathcal{F}_\text{LR}=\SI{98.5(0.3)}{\%}$ in \SI{310}{\nano\second}.
Similarly to the reset protocol, the leakage recovery can be extended to higher levels of the qubit, if necessary.
Furthermore, the leakage recovery operation does not have to be applied every Clifford sequence to dynamically account for the trade-off between leakage and decoherence rates, as discussed in Appendix~\ref{app:LR_ncl}.
A reduction of the required drive length for both the reset and the leakage recovery can be achieved by using a resonator with a stronger coupling to the tunable coupler and to the environment, lowering the time needed for the residual excitation in the resonator to decay.
A reduction in the operation length would also reduce the impact of qubit decoherence on the fidelity of the leakage recovery, as seen from Eq.~\eqref{eq:error_LR}.
Increasing the operation speed further can in principle be achieved with stronger pulses.
However, for strong flux pulses higher-order effects during the drive appear and lead to a degradation in operation fidelity, as predicted by~\cite{Petrescu2023, Lagemann2022}. 
A detailed investigation of these limits, together with possible mitigation techniques is of high interest, but beyond the scope of this work and will be subject to forthcoming studies.

We utilize the parametric drive to control a state-dependent resonator shift $\chi/2\pi$ tunable between \SI{-1.25}{\mega\hertz} and \SI{1.25}{\mega\hertz}, almost three orders of magnitude larger than the negligible static contribution of $|\chi_\text{static}|/2\pi<\SI{4}{\kilo\hertz}$.
We set the parametrically activated shift to $\chi/2\pi = \SI{-0.53}{\MHz}$ and implement a coupler-driven parametric readout with single-shot assignment fidelity of \SI{88(0.4)}{\%}, mostly limited by the qubit decay rate.
We, thereby, explicitly show that a parametrically induced single-shot readout of a superconducting qubit can be reached, going beyond the recent implementation in~\cite{Noh2023}.
Improving the parametric readout fidelity can be achieved by increasing the direct coupling between resonator and coupler, leading to a larger effective $\chi$-shift and greater separation of the measurement signals.
Similarly, the use of quantum-limited amplifiers combined with improved readout pulse shaping~\cite{Jerger2024, Chen2023, McClure2016} will lead to an enhanced single-shot fidelity in future experiments.
Lastly, we realize a controlled-Z gate between neighbouring qubits with a fidelity of $\mathcal{F}_\text{CZ}=\SI{98.3(0.23)}{\%}$ within \SI{339}{\nano\second}, completing this set of elementary operations.
The fidelity of the implemented two-qubit gate is primarily limited by decoherence and can be improved further by increasing qubit coherence times and applying pulse optimization techniques.

An architecture utilizing the demonstrated methods presents numerous advantages and possible extensions.
Utilizing fixed-frequency qubits and a tunable element to mediate couplings allows for large on/off ratios and protects the qubits from dephasing due to flux noise and from resonator-photon induced decoherence.
This approach complements alternative decoupling strategies, e.g. multi-mode qubits~\cite{Gambetta2011, Gyenis2021, Pfeiffer2024}.
Additionally, two-qubit gates are not limited to the demonstrated controlled-Z gate but can be extended to include iSWAP and bSWAP gates~\cite{McKay2017, Roth2017, Ganzhorn2020}, or to realize long-distance state-transfer protocols that require simultaneous controllable interactions between neighbouring qubits along a chain~\cite{Gu2021, Warren2023, Naegele2022, Roy2024}.
Since couplings can be selectively activated, more than two qubits can be connected to the same coupler without necessarily introducing unwanted qubit-qubit crosstalk, as discussed in Appendix~\ref{app:ZZ_crosstalk}.
Given proper mitigation of frequency crowding, multi-qubit couplers increase device connectivity and can be used to implement multi-qubit operations beyond two qubits ~\cite{Glaser2023, Kim2022, Zhang2022, Simakov2023}.
Furthermore, the fact that multiple qubits are coupled to the same resonator not only reduces the total number of resonators but also enables the application of advanced readout techniques.
For example, dynamic control over the $\chi$-shifts enables the simultaneous single-shot readout of multiple qubits~\cite{Filipp2009, Reed2010}.
Likewise, $\chi$-shifts of equal magnitude can be used to perform direct parity measurements~\cite{Royer2018, Noh2023}, highly relevant for quantum error correction in future large-scale quantum processors.
\section{Acknowledgments}
We thank Alexandru Petrescu and Ioan Pop for insightful discussions and helpful comments.
This work received financial support by the European Union's Horizon 2020 research and innovation programme 'MOlecular Quantum Simulations' (MOQS; Grant-Nr. 955479), the EU MSCA Cofund 'International, interdisciplinary and intersectoral doctoral programme in Quantum Science and Technologies' (QUSTEC; Grant-Nr. 847471), the BMBF programs 'German Quantum Computer based on Superconducting Qubits' (GeQCoS; No 13N15680) and the MUNIQC-SC initiative (No 13N16188) as well as the German Research Foundation via the projects 445948657 and Germany's Excellence Strategy "EXC-2111-390814868" and the Munich Quantum Valley, which is supported by the Bavarian state government with funds from the Hightech Agenda Bayern Plus.

\begin{appendices}

\section{\texorpdfstring{\\* \vspace{2mm}}~Effective parametric coupling strength}
\label{app:eff_ham}
The modulation of the coupler frequency with a flux pulse ${\varphi_\text{ext}(t) =\varphi_\text{dc} + a_\text{D} \sin(\omega_\text{D}t)}$ at the drive frequency $\omega_\text{D}$ leads to an effective coupling between two states when the coupler is driven at a harmonic of their frequency difference.
Here, $\varphi_\text{dc}$ is the static flux offset and $a_\text{D}$ the drive amplitude.
To derive the effective coupling under this parametric drive, as given in Eq.~\eqref{eq:g_param}, we write the coupler frequency $\omega_\text{C}$ as a Taylor series around the static flux offset
\begin{equation}
\label{eq:taylor_app}
    \omega_\text{C} = \sum_{n=0}^\infty \frac{1}{n!} \frac{\partial^n 
    \omega_\text{C}}{{\partial \varphi}^n}\big|_{\varphi_\text{dc}} a_\text{D}^n \sin^n (\omega_\text{D}t).
\end{equation}
To remove the explicit time dependency of the coupler, we first use the identities
\small
\begin{equation}
\begin{aligned}
&\sin ^{2 n} (x) =\frac{1}{2^{2 n}}\left(\begin{array}{c}
2 n \\
n
\end{array}\right)+\sum_{k=0}^{n-1}\frac{(-1)^{n+k}}{2^{2 n-1}}\left(\begin{array}{c}
2 n \\
k
\end{array}\right) \cos [2(n-k) x] \\
&\sin ^{2 n+1} (x) =\sum_{k=0}^n\frac{(-1)^{n+k}}{4^n}\left(\begin{array}{c}
2 n+1 \\
k
\end{array}\right) \sin [(2 n+1-2 k) x],
\end{aligned}
\end{equation}
\normalsize
to expand $\sin^n (\omega_\text{D}t)$.
Reordering the sum in Eq.~\eqref{eq:taylor_app} results in
\begin{equation}
\omega_\text{C}(t) = \overline{\omega}_\text{C}+\sum_{m=1}^\infty D_m\cos\qty[m\qty(\omega_\text{D}t-\frac{\pi}{2})]
\end{equation}
as terms of a Fourier series.
Here, $\overline{\omega}_\text{C}$ is the average coupler frequency $\overline{\omega}_\text{C}$ during drive modulation and the Fourier coefficients $D_m$ take the form
\footnotesize	
\begin{equation}
\label{eq:Fourier_coeffs}
\begin{aligned}
\overline{\omega}_C &= \sum_{n=0}^\infty \frac{\partial^{2n} 
    \omega_\text{C}}{{\partial \varphi}^{2n}}\Big|_{\varphi_\text{dc}} \frac{ a_D^{2n}}{4^n (n!)^2}&\\
    D_m &= \sum_{n=\frac{m}{2}}^\infty  \frac{\partial^{2n} 
\omega_\text{C}}{{\partial \varphi}^{2n}}\Big|_{\varphi_\text{dc}} \frac{2 a_D^{2n}}{4^{n}\qty(\frac{2n-m}{2})!\qty(\frac{2n+m}{2})!}&\text{$m$ even}\\
    D_m &= \sum_{n=\frac{m-1}{2}}^\infty  \frac{\partial^{2n+1}
\omega_\text{C}}{{\partial \varphi}^{2n+1}}\Big|_{\varphi_\text{dc}} \frac{ a_D^{2n+1}}{4^{n}\qty(\frac{2n+1-m}{2})!\qty(\frac{2n+1+m}{2})!}&\text{$m$ odd.}\\
\end{aligned}
\end{equation}
\normalsize
The resulting modulation of the system in the manifold of driven states ($\ket{\text{A}},\ket{\text{B}}$) and coupler ($\ket{\text{C}}$) is then described by the Hamiltonian
\begin{equation}  
    \label{eq:Hamilton_floquet_}
		\begin{split}
			\hat{H}(t)/\hbar   &= k \omega_\text{D} \ket{\text{B}}\bra{\text{B}} + \Delta_\text{C} \ket{\text{C}}\bra{\text{C}}  \\
             &+ \sum_{\substack{i,j \in\text{\{A,B,C\}} \\i \neq j}} g_{ij}\ket{i}\bra{j}\\
             &+\sum_{m=1}^\infty D_m\cos\qty[m\qty(\omega_\text{D}t-\frac{\pi}{2})] \ket{\text{C}}\bra{\text{C}},\\
        \end{split}
\end{equation}
where $\omega_\text{D} = (\omega_\text{B}-\omega_\text{A})/k$ the drive frequency, {$\Delta_\text{C} = \overline{\omega}_\text{C}-\omega_\text{A}$} the shifted average coupler frequency during the drive and $g_{ij}$ the static couplings between the states, where we have shifted the energies by $-\omega_\text{A}$.
To derive the effective Hamiltonian under parametric drive we use the periodicity of the Hamiltonian $\hat{H}(t) =\hat{H}(t+2\pi/\omega_\text{D})$ and apply a Floquet-transformation~\cite{Schweizer2019, Petrescu2023} to remove the time dependency due to the coupler modulation.
To do so we apply the transformation $R = e^{S}$, where
\begin{equation}
    \label{eq:S}
    \begin{aligned}
        S &= i k\omega_\text{D} t \ket{\text{B}}\bra{\text{B}}\\
        & +i \sum_{m=1}^\infty \frac{D_m}{m \omega_\text{D}}\sin\qty[m\qty(\omega_\text{D}t-\frac{\pi}{2})]\ket{\text{C}}\bra{\text{C}}.
    \end{aligned}
\end{equation}
This allows to rewrite the Hamiltonian as 
\begin{equation}
    \label{eq:H'}
    \begin{aligned}
        \hat{H}'(t)&=R \hat{H} R^{\dagger}-i R \partial_t R^{\dagger}\\
        &= \Delta_\text{C}\ket{C}\bra{C}+ g_{AB} e^{-i k\omega_\text{D} t}\ket{A}\bra{B} \\
        &+ g_{AC} \Lambda  \ket{A}\bra{C} + g_{BC} \Lambda e^{ik\omega_\text{D} t} \ket{B} \bra{C} + \text{h.c.}
    \end{aligned}
\end{equation}
Here, we have used the Jacobi-Anger expansion to expand the Fourier series in the exponentials and introduced 
\begin{equation}
    \label{eq:series}
    \Lambda = \prod_{m=1}^\infty \sum_{n=-\infty}^{\infty} (-i)^{2n+nm} J_n\qty(\frac{D_m}{m\omega_\text{D}})e^{i n m \omega_\text{D}t},
\end{equation}
where $J_n$ are the Bessel functions of the first kind of $n$-th order.
Collecting terms with the same periodicity gives
\begin{equation}
\hat{H}'(t)=\sum^{\infty}_{l=-\infty} \hat{H}^{(l)} e^{i l \omega_\text{D} t},
\end{equation}
where $\hat{H}^{(l)}$ are time-independent and are used to obtain the Floquet Hamiltonian to lowest order
\begin{equation}
\label{eq:floquet_H}
\hat{H}_F=\hat{H}^{(0)}+\sum_{l>0}^{\infty} \frac{1}{l \omega_\text{D}}\left[\hat{H}^{(+l)}, \hat{H}^{(-l)}\right]+\mathcal{O}\left(\frac{1}{\omega_\text{D}^2}\right).
\end{equation}
A general expression for $\hat{H}^{(l)}$ includes all terms in Eq.~\eqref{eq:H'} rotating at the frequency $l \omega_\text{D} t$.
However, since ${J_n(x\rightarrow0)\approx x^\abs{n}/2^\abs{n}\abs{n}!}$ one can truncate the series in Eq.~\eqref{eq:series} to a suitable choice of Bessel function order.
To find the largest contribution to the effective coupling of $\ket{A}$ and $\ket{B}$ in the frame of the drive for arbitrary drive harmonics, we keep only terms in Eq.~\eqref{eq:series} that are a product of at most one Bessel functions of order {$n \ne 0$}. 
This results in
\begin{equation}
\tilde{g}_{\text{A}\text{B}} \approx -g_{\text{A}\text{C}}g_{\text{B}\text{C}} \frac{(-i)^k}{k\omega_\text{D}} J_1\qty(\frac{D_k}{k\omega_\text{D}})\propto a_\text{D}^k, 
\end{equation}
where we assumed $J_0(D_m/m\omega_\text{D})\approx 1$, for all $m\ne k$, giving the value in Eq.~\eqref{eq:g_param}. Here, the proportionality to the drive amplitude is found by truncating Eq.~\eqref{eq:Fourier_coeffs} to first order and inserting for $D_k$.

The general Floquet Hamiltonian is written as
\begin{equation}  
    \label{eq:floquet_frame_drive}
		\begin{split}
			\hat{H}(t)/\hbar   &= \tilde{\omega}_\text{A} \ket{\text{A}}\bra{\text{A}} + \tilde{\omega}_\text{B} \ket{\text{B}}\bra{\text{B}} + \tilde{\Delta}_\text{C} \ket{\text{C}}\bra{\text{C}}  \\
             &+ \sum_{\substack{i,j \in\text{\{A,B,C\}} \\i \neq j}} \tilde{g}_{ij}\ket{i}\bra{j},\\
        \end{split}
\end{equation}
with the parametric frequency shifts $\tilde{\omega}_\text{A}$, $\tilde{\omega}_\text{B}$, $\tilde{\Delta}_\text{C}$ and couplings $\tilde{g}_{ij}$. 
We find expressions for these parameters in the case of $k=2$, by truncating the Fourier series to second order, by only including Bessel functions up to second order and by only taking the first two terms in the sum of Eq.~\eqref{eq:floquet_H}
\begin{widetext}
\begin{equation}
\begin{gathered}
    \tilde{\Delta}_\text{C}  =\Delta_\text{C}-\tilde{\omega}_\text{A}-\tilde{\omega}_\text{B} \\
    \tilde{g}_{\text{A}\text{C}} =  g_{\text{A}\text{C}}J_{0,0}+\frac{g_{\text{A}\text{B}}g_{\text{B}\text{C}}}{2\omega_\text{D}} \qty(J_{0,2}+J_{2,1} )\\
        \tilde{\omega}_\text{A} =\frac{1}{\omega_\text{D}}\qty(\frac{g_{\text{A}\text{B}}^2}{2}+g_{\text{A}\text{C}}^2\qty[4J_{1,0}J_{1,1}-2\qty{J_{0,1}J_{2,2}+J_{0,0}J_{2,1}}]) \\
    \tilde{g}_{\text{B}\text{C}}  =  -g_{\text{B}\text{C}}\qty(J_{0,1}+J_{2,0}+J_{2,2})+\frac{g_{\text{A}\text{B}}g_{\text{A}\text{C}}}{2\omega_\text{D}} \qty(-J_{0,1}+J_{2,0}+J_{2,2} )\\
    \tilde{\omega}_\text{B} =\frac{1}{2\omega_\text{D}}\qty(g_{\text{C}\text{B}} ^ { 2 } \qty[J_{0,0}^2-J_{0,2}^2-J_{2,1}^2+2\qty{J_{1,0}^2-J_{1,2}^2-J_{0,1}J_{2,2}}+4 J_{1,1}\qty{J_{1,2}-J_{1,0}}] -g_{\text{A}\text{B}}^2)\\
    \tilde{g}_{\text{A}\text{B}}   = \frac{ g_{\text{A}\text{C}}g_{\text{B}\text{C}}}{2\omega_\text{D}} \qty(J_{0,0}\qty[J_{0,1}-J_{2,0}]+J_{0,1}\qty[J_{0,2}+J_{2,1}] +J_{0,2}J_{2,2}+J_{2,0}J_{2,1}+2 J_{1,0}\qty[J_{1,0}+J_{1,1}-J_{1,2}]+2 J_{1,1} J_{1,2}-4 J_{1,1}^2  ) \\
\end{gathered}
\end{equation}
\end{widetext}
where we introduced $J_{n,m} = J_n(D_1/\omega_\text{D}) J_m(-D_2/2\omega_\text{D})$.
The higher order effect of the additional non-zero parametric coupling to the coupler in the frame of the drive can be estimated using an additional Schrieffer-Wolff transformation of the Floquet Hamiltonian to remove the coupler.
The transformation is valid when the coupler is detuned from nearby transitions and the detuning is larger than the couplings ($\tilde{\Delta}_\text{C}\gg \tilde{\omega}_\text{A},\tilde{\omega}_\text{B}, \tilde{g}_\text{ij}$). 
The correction to the direct effective coupling between $\ket{A}$ and $\ket{B}$ is then
\begin{equation}
\tilde{g}'_{\text{A}\text{B}}\approx \tilde{g}_{\text{A}\text{B}}-2 \frac{\tilde{g}_{\text{A}\text{C}} \tilde{g}_{\text{C}\text{B}}}{\tilde{\Delta}_\text{C}},
\end{equation}
reminiscent of the simplified effective coupling expression found in the case of static couplings, where the detuning and couplings have been replaced by the parametric ones~\cite{Sete2021floating}.


\section{\texorpdfstring{\\* \vspace{2mm}}~Experimental setup}
\label{app:exp_setup}
\begin{figure}[t!]
\includegraphics[width=0.48\textwidth]{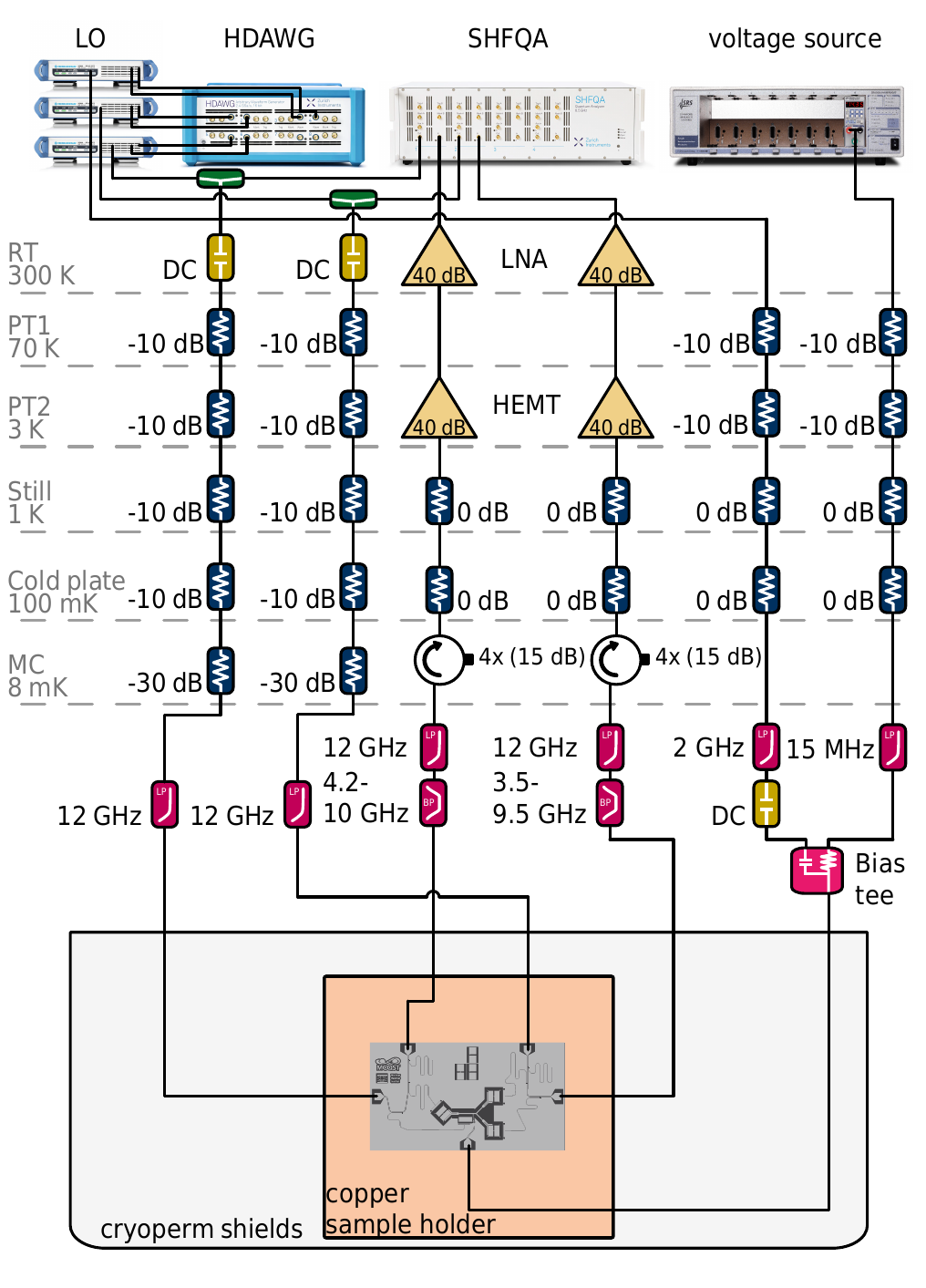} 
\caption{Wiring and control electronics of the experiment. See main text for details.\label{fig:fridge_setup}}
\end{figure}
\begin{figure}[b]
\includegraphics{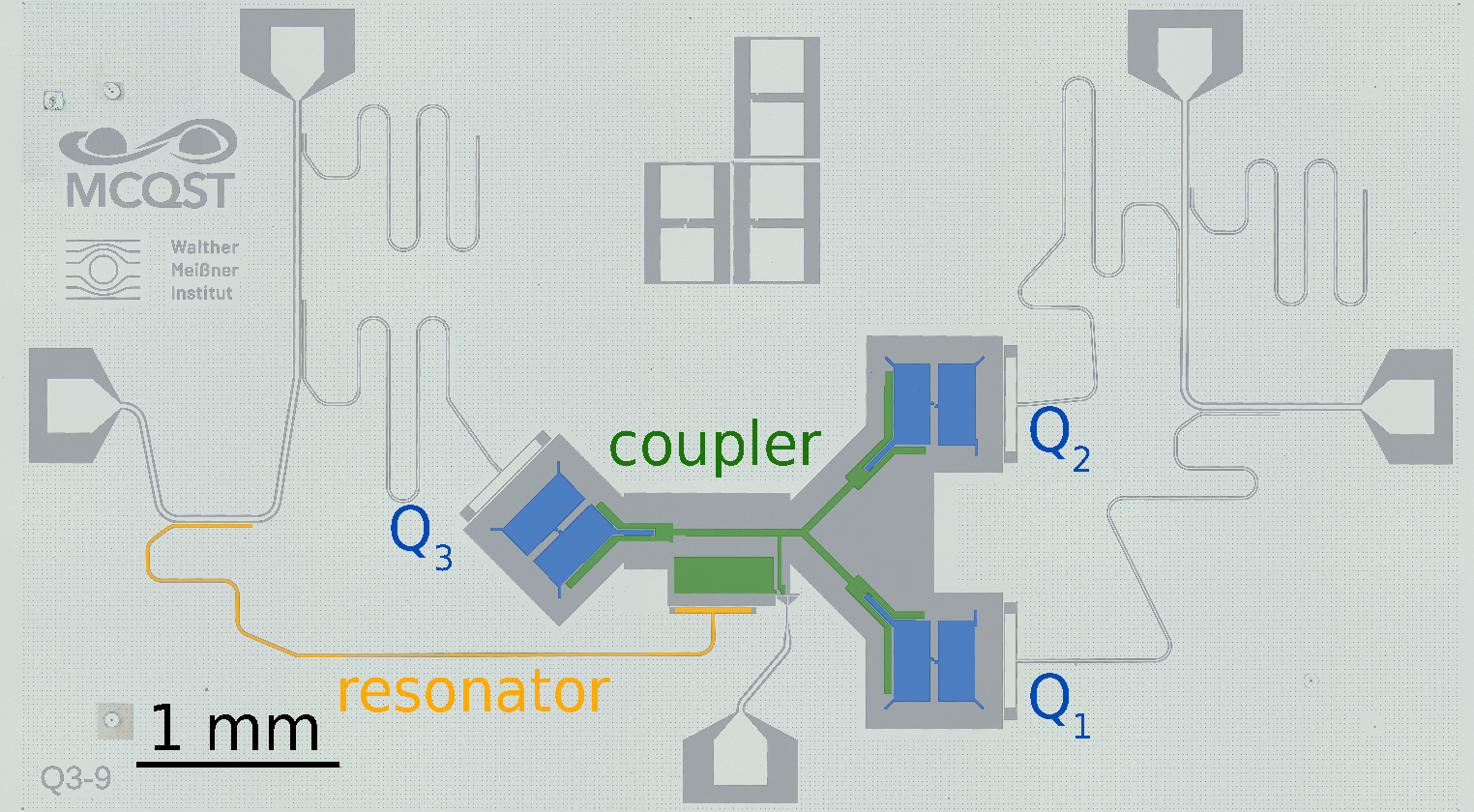} 
\caption{Microscope image of the chip used in this work. The qubits (blue), coupler (green) and resonator (yellow) are highlighted. Single-qubit drives are applied through the feedline. Qubit measurements are performed using dispersively coupled resonators connected to feedlines, unless stated otherwise.
\label{fig:sample} }
\end{figure}
\begin{figure}[b]
\includegraphics{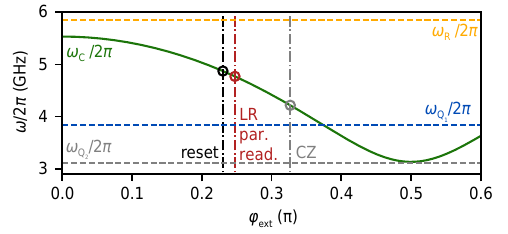} 
\caption{
    \label{app:wC_flux}
    Coupler frequency (green) as a function of external flux $\varphi_\text{ext}$.
    Horizontal dashed lines show the frequencies of the resonator (yellow), $\text{Q}_1$ (blue) and $\text{Q}_2$ (grey) against external flux.
    The operating points for the controlled-Z gate (grey), reset (black), LR and parametric readout (red) are indicated by the circles and vertical dashed-dotted lines.
    }
\end{figure}
\begin{table*}
  \begin{tabular}{ccc}
        \begin{tabular}{ | c | c | c | c | c |} 
         \hline
           & $\text{Q}_1$ & $\text{Q}_2$ & \text{C} & \text{R}  \\ \hline
          $\omega_i / 2\pi$ & \SI{3.83}{\giga\hertz} & \SI{3.11}{\GHz}& \SI{3.13}{}-\SI{5.45}{\giga\hertz} &\SI{5.85}{\giga\hertz} \\ 
          $\alpha_i / 2\pi$ & \SI{-205}{\mega\hertz}& \SI{-216}{\mega\hertz}& \SI{-161}{\mega\hertz}& \\ 
           $\Gamma_1 / 2\pi$ & \SI{4.2(1.6)}{\kHz}& \SI{4.7(1.0)}{\kHz}& \SI{20(2)}{\kHz}&\SI{770(3)}{\kHz} \\ 
         $\Gamma_\phi / 2\pi$ & \SI{4.4(1.3)}{\kHz}& \SI{11.3(8.0)}{\kHz}& \SI{143(10)}{\kHz}&  \\
         \hline
        \end{tabular}
        & \hspace{-2.5mm} &
        \begin{tabular}{ | c | c | c | c | c |} 
         \hline
           $g_{ij}/ 2\pi$ & $\text{Q}_1$ & $\text{Q}_2$ & \text{C} & \text{R}  \\   \hline
           $\text{Q}_1$ & & \SI{15}{\mega\hertz} &\SI{115}{\mega\hertz} &\SI{9}{\mega\hertz}\\ 
           $\text{Q}_2$ & \SI{15}{\mega\hertz} &  &\SI{110}{\mega\hertz} &\SI{5}{\mega\hertz} \\
            C & \SI{115}{\mega\hertz} & \SI{110}{\mega\hertz} & &\SI{-75}{\mega\hertz} \\ 
          R & \SI{9}{\mega\hertz} & \SI{5}{\mega\hertz} & \SI{-75}{\mega\hertz} & \\ 
           \hline
        \end{tabular}
    \end{tabular}
    \caption{
    Device parameters of the qubits (Q$_1$, Q$_2$), coupler (C) and resonator (R). 
    The frequencies $\omega_i$, anharmonicities $\alpha_i$ and static couplings $g_{ij}$ between elements are given for zero flux offset.
    The decay rates were measured when the coupler is placed at the operating point for the LR and parametric readout protocol.
    The static crosstalk between Q$_1$ and Q$_2$ was measured to be $\zeta_{1,2} = \SI{-301.5(2.4)}{\kHz}$ at the operating point for the two-qubit gate.
    All parameters are determined using standard characterization methods. \label{tab:params} 
    }
\end{table*}

A schematic of the experimental setup is shown in Fig~\ref{fig:fridge_setup}.
The qubit control pulses are generated using two channels of an arbitrary waveform generator (AWG) from Zurich Instruments (HDAWG750MHz) and a local oscillator (LO) from Rohde\&Schwarz (SGS100A-SGMA), using a sideband of
\SI{100}{\MHz}.
The measurement pulses are generated using a Zurich Instruments ‘Super High Frequency Quantum Analyzer' (SHFQA). 
We do not use charge-coupled direct drive lines on the chip for qubit control.
Instead, the output of the LO for single-qubit control is combined with the output line of the SHFQA at room temperature using a commercial combiner (Mini-Circuits ZX10-2-71+).
The signal line is filtered with a DC-block at room temperature, attenuated by \SI{-70}{\dB} distributed over the different temperature stages and filtered with a \SI{12}{\GHz} low pass filter.
We use two $4-\SI{12}{\GHz}$ isolators on each output line and \SI{0}{\dB} attenuators for thermal anchoring. 
The output signal is amplified by a $4-\SI{8}{\GHz}$ \SI{40}{\dB} HEMT cryogenic low-noise amplifier (LNF-LNC4\_8F) thermalized at the \SI{3}{\kelvin} stage.
The signal is further amplified by a \SI{40}{\dB} low noise room temperature amplifier (BZ-04000800-081045-152020), before being digitized at the SHFQA input.
The {AC-component} of the coupler is generated using two AWG channels and a LO placed at \SI{1}{\GHz}. The flux line is attenuated by \SI{-20}{\dB} and filtered using a \SI{2}{\GHz} low-pass filter and a DC-block on the mixing chamber stage.
The DC-bias is created using a Stanford Research Systems voltage source (SRS-SIM928) and attenuated by \SI{-20}{\dB}, before being filtered using a \SI{15}{\MHz} low-pass filter.
The AC- and DC-lines are combined at the mixing chamber using a Mini-circuits bias tee (VHF-3100+). 

The chip, shown in Fig.~\ref{fig:sample}, contains three qubits ($\text{Q}_1, \text{Q}_2, \text{Q}_3$), all connected to a single flux-tunable coupler.
All qubits and the coupler are capacitively coupled to an individual resonator individually.
Two resonators each are connected to either the right or the left feedline on the chip.
The flux through the SQUID-loop of the coupler is generated using a grounded flux line.
The chip is housed in a five-port box, with four ports being used for the two on-chip feedlines and one port being used for the coupler flux line.
The parameters of the qubits $\text{Q}_1$ and $\text{Q}_2$, the coupler and resonator have been determined using standard characterization methods and are listed in Table~\ref{tab:params}.
For the experiments in this work $\text{Q}_3$ is not used.
We determine the decay rate $\Gamma_{f\rightarrow e} = \SI{6.3(0.7)}{\kHz}$ of $\text{Q}_1$ for the leakage recovery section by preparing the qubit in $\ket{f}$ and monitoring the evolution of the population in the first three states.
The uncoupled frequencies of the elements used in this work are shown in Fig.~\ref{app:wC_flux}, the flux bias of the coupler used for implementing the different protocols is indicated by the vertical lines.
The flux bias is chosen to maximize the effective coupling strength while ensuring that no leakage from other transitions is observed.
Performing all operations in the same sequence can be done with the use of baseband pulses to move the coupler to the desired operation points during the parametric pulse, as for example implemented in~\cite{Chu2023}.

Single-qubit gates are realized by \SI{30}{\nano\second}-long pulses with a DRAG-Gaussian~\cite{Motzoi2009} envelope for the reset, parametric readout and two-qubit gate experiments.
To reduce the baseline leakage to higher-excited qubit states, we have increased the single-qubit pulse length during the leakage randomized benchmarking sequences to \SI{100}{\nano\second}.
The parametric drive pulses used for reset, LR, parametric readout and two-qubit gate have a flat-top Gaussian envelope with a $\sigma$-width of the rise- and fall-time of \SI{10}{\nano\second}, \SI{1}{\nano\second}, \SI{10}{\nano\second} and \SI{10}{\nano\second}, respectively.

\section{\texorpdfstring{\\* \vspace{2mm}}~State dynamics during decay} 
\label{app:dynamic_decay}
By driving the transition $\ket{e,0}\leftrightarrow\ket{g,1}$ using a parametric pulse, we activate an effective coupling $\tilde{g}_\text{QR}$ in the frame of the drive, resulting in the effective Hamiltonian
\begin{equation} 
\label{eq:basic_H_RWA}
    \begin{split}
	\hat{H}_\text{eff}/\hbar = \tilde{g}_\text{QR}\ket{\text{e,0}}\bra{\text{g,1}}+\text{h.c.}
	\end{split}
\end{equation}
To derive an analytic expression for the qubit-resonator dynamics during this parametric coupling we consider only the single excitation manifold and model qubit and resonator decay by introducing non-Hermitian terms in the Hamiltonian~\cite{Magnard2018, Zhou2021}
\begin{align}  
    \hat{H}_\text{eff}^\prime/\hbar = 
    \begin{pmatrix}
    -i\Gamma_1/2 & \tilde{g}_\text{QR}\\[6pt]
    \tilde{g}_\text{QR}^* & -i\kappa_\text{R}/2\\[6pt]
    \end{pmatrix}.
\end{align}	
Here, $\Gamma_1$ is the decay rate of the qubit and $\kappa_\text{R}$ is the decay rate of the resonator.
We distinguish three different
\begin{figure}[H]
    \includegraphics{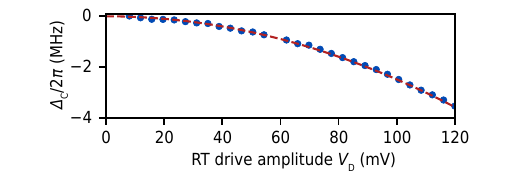}
    \caption{
        \label{app:flux_ampl_calibration} 
        Frequency shift $\Delta_\text{C}$ of the coupler against applied drive amplitude $V_\text{D}$.
        The red dashed line shows the fit to the data to extract the conversion factor between room temperature amplitude $V_\text{D}$ and the effective flux drive amplitude $a_\text{D}$ on the chip.
    }
\end{figure}
\hspace{-4mm}regimes of system dynamics
\begin{align} 
\label{eq:decay_dynam}
    &\hspace{0mm} P_\text{e}(t) = \qty|\bra{\text{e,0}}e^{-i \hat{H}_\text{eff}^\prime t/\hbar} \ket{\text{e,0}}|^2 =\\
    &\hspace{2mm} = e^{-\frac{\kappa_\Sigma}{2}t}
    \begin{cases}
       \qty[\cosh(\text{M}t)+\frac{\kappa_\Delta}{4\text{M}}\sinh(\text{M}t)]^2 & \qty|\tilde{g}_\text{QR}|<\frac{\kappa_\Delta}{4} \\
        \qty(1 + \frac{\kappa_\Delta}{4}t)^2 & \qty|\tilde{g}_\text{QR}|=\frac{\kappa_\Delta}{4}\\
        \qty[\cos(\text{M}t)+\frac{\kappa_\Delta}{4\text{M}}\sin(\text{M}t)]^2  & \qty|\tilde{g}_\text{QR}|>\frac{\kappa_\Delta}{4}, \\
   \end{cases}
   \notag
\end{align}
where $\kappa_{\Sigma/\Delta} = \kappa_\text{R}\pm\Gamma_1$ and $\text{M} =\qty|\sqrt{\qty|\tilde{g}_\text{QR}|^2-\qty(\frac{\kappa_\Delta}{4})^2}|$.
For weak coupling the system is overdamped and no swaps between the states occur during the decay.
When $\qty|\tilde{g}_\text{QR}|=\kappa_\Delta/2$ the system is critically damped and exactly one swap occurs before reaching equilibrium.
For larger couplings the system is underdamped and reaches equilibrium after multiple swaps between the states.
When using the Eq.\eqref{eq:decay_dynam} for fitting experimental data, we include a scaling factor to
\begin{equation}  
	\gamma(\tau) = \int_0^\tau \text{A}_\text{D}(t) dt,
\end{equation}
to account for the envelope of the parametric pulse $\text{A}_\text{D}(t)$ with a total duration $\tau$.
\section{\texorpdfstring{\\* \vspace{2mm}}~Flux pulse amplitude calibration}
\label{app:flux_pulse_calibration}

The magnetic flux $\varphi_\text{ext}$ generated by the room temperature drive $V(t)=V_\text{D}\sin(\omega_D t)$ is calibrated by measuring the shift $\Delta_\text{C} = \overline{\omega}_\text{C}-\omega_\text{C}$ of the average coupler frequency during parametric drive compared to the static coupler frequency, similar to the calibration in~\cite{Sete2021parametric}.
We vary the amplitude $V_\text{D}$ of the device output and measure the effective shift in the coupler frequency with a Ramsey experiment, shown in Fig.~\ref{app:flux_ampl_calibration}.
We fit this shift with a numerical model of the coupler frequency flux dependency to extract the magnitude $a_\text{D}$ of the magnetic flux drive.

\section{\texorpdfstring{\\* \vspace{2mm}}~Calibration and QPT of LR operation}
\label{app:LR_cal}
\begin{figure}[t]
\includegraphics{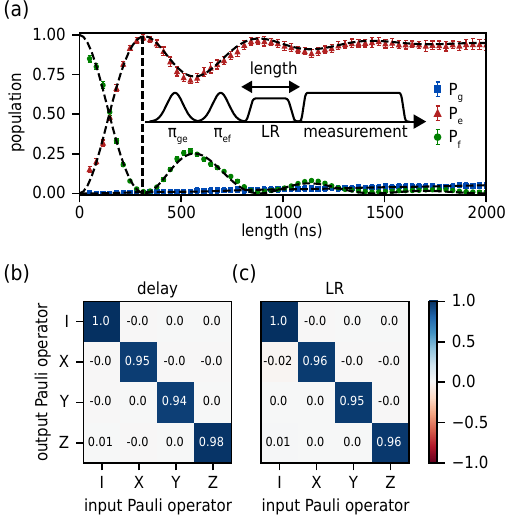}
\caption{
Calibration and characterization of LR. 
(a) Qubit dynamics during the LR drive showing the ground state (blue squares), the first-excited state (red triangles) and second-excited state (green circles). 
Dashed lines represent a fit to the experimental data, see main text for details.
(b-c) Pauli transfer matrices as obtained from Quantum process Tomography when applying either a delay (b) or the LR operation (c).
\label{fig:LR_calib}}
\end{figure}

To calibrate the LR operation we first prepare the second-excited state of the qubit using a $\pi_{ge}$ pulse on the $\ket{g} \leftrightarrow\ket{e}$ transition followed by a $\pi_{ef}$ pulse on the $\ket{e}\leftrightarrow\ket{f}$ transition.
Next, we drive the coupler at the second-harmonic ($k=2$) of the $\ket{f,0}\leftrightarrow \ket{e,1}$ transition with varying drive length and measure the population in the first three qubit states.
The dynamics are plotted as a function of the parametric drive length in Fig.~\ref{fig:LR_calib}(a).
The second-excited state decays with oscillations to the first-excited state, which in turn decays to the ground state.
To account for this additional decay, we modify the model for the two state dynamics in Eq.~\eqref{eq:decay_dynam}, by multiplying the population in $\ket{e}$ with the prefactor $e^{-t\Gamma_1}$ and by approximating the population in the ground state as $P_g \approx (1-e^{-t\Gamma_1}) P_e$.
The model agrees well with the data, with the effective coupling strength $\tilde{g}_{\text{Q}\text{R}}$ as the only free fit parameter.
From the fit we find a full swap occurring after \SI{310}{\nano\second}.
During the parametric flux drive the qubit frequency is shifted, leading to an additional phase on the qubit subspace \{$\ket{g}, \ket{e}$\}.
To remove this phase, we calibrate an additional virtual-Z gate and apply it after the LR operation.\\
We quantify the decoherence introduced by the LR operation by performing Quantum Process Tomography (QPT)~\cite{Nielsen2010} on the computational subspace of the qubit when applying either a delay of equal length as the LR operation, or the LR operation itself.
The resulting Pauli transfer matrices, shown in Fig.~\ref{fig:LR}(b,c), give fidelities of \SI{98.0(1.1)}{\%}, \SI{97.9(0.9)}{\%} respectively, suggesting that the decoherence during the LR operation drive is not significantly increased.

\section{\texorpdfstring{\\* \vspace{2mm}}~Readout calibration}
\label{app:gef_readout}
\begin{figure}[t]
\includegraphics{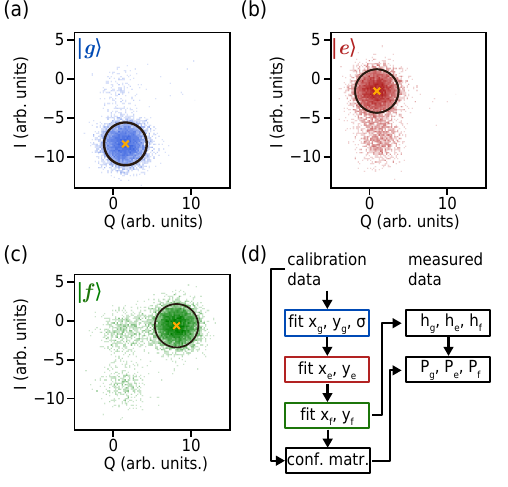} 
\caption{Calibration sequence for the processing of measurement data using 2D Gaussian distributions. (a) Initial fit of the ground state distribution from which the center of the Gaussian and width are extracted. (b)/(c) Fit of the first/second state using the same width extracted from the fit of the ground state population. (d) Schematic representation of the calibration procedure for the determination of the population in the first three qubit states, see text for a detailed description.\label{fig:calib_gef}
}
\end{figure}

Here, we develop a method for the accurate estimation of the population in the first three qubit states from single-shot data.
The method relies on a sequential fit of single-shot data of the first three qubit states used for calibration, see Fig.~\ref{fig:calib_gef}.
First, a two dimensional Gaussian distribution 
\begin{equation}
   f(x,y) = \text{h} \cdot e^{-\frac{(x-x_0)^2+(y-y_0)^2}{2\sigma^2}}
\end{equation}
is fitted to a 2D histogram of the single-shot measurement of the ground state, where $\text{h}$ denotes the height, $\sigma$ the width and $(x_0,y_0)$ the center of the distribution.
From this fit the width and center of the ground state are extracted, Fig.~\ref{fig:calib_gef}(a). 
Next, we fit the position of the center of the first-excited qubit-state with a Gaussian distribution, where the width is given by the fit result of the ground state, Fig.~\ref{fig:calib_gef}(b).
We repeat the previous procedure for the second-excited state, Fig.~\ref{fig:calib_gef}(c).

Lastly, the same three data-sets are fitted each with a sum of the three Gaussian distributions
\begin{equation}
    F(x,y) = \sum_{i \in \{e,g,f\}} \text{h}_i\cdot e^{-\frac{(x-x_i)^2+(y-y_i)^2}{2 \sigma^2}},
    \label{eq:gauss_distr}
\end{equation}
where the width $\sigma$ and the centers $(x_i,y_i)$ are set to the previously obtained values.
From these fits the height of each of the three distributions is extracted for each data-set and used to create a confusion matrix.
All measured data is then fitted using the same distribution from Eq.~\eqref{eq:gauss_distr}.
The inverse of the confusion matrix is applied to estimate the population in the first three qubit states.
The full workflow of the fitting routine and data extraction is shown in Fig.~\ref{fig:calib_gef}(d).
We find that our method is robust against a fluctuating qubit decay rate and exhibits less variance compared to support vector machines, in particular, when the qubit states are not well separated in phase space.

\section{\texorpdfstring{\\* \vspace{2mm}}~Effect of Leakage on RB}
\label{app:rates}
To characterize the performance of the leakage recovery (LR) operation we utilize randomized benchmarking (RB) sequences of length $\text{n}_\text{Cl}$ with interleaved leakage injection and LR operations.
Here, we derive the expressions for the equilibrium leakage $A_2$, i.e. the population $P_f$ in the second excited state $\ket{f}$ in the limit $\text{n}_\text{Cl}\rightarrow\infty$, and for the average error $\varepsilon$ in the RB experiments of Fig.~\ref{fig:LR}.

For the estimation of $A_2$ we describe the average qubit population per Clifford gate with the vector
\begin{equation}
\vec{P} = 
\begin{bmatrix}
    P_{\text{sub}} \\
    P_{f}   \\
    P_{\text{R}}
\end{bmatrix},
\end{equation}
where $P_{\text{sub}}$ is the population in the computational subspace, $ P_{f}$ is the qubit population in the $\ket{f}$-state and $P_{\text{R}}$ is the population in the resonator.
Here, we assume no coherence between the subspace population, leaked population and the population in the resonator.

The leakage injection operation transfers population from $\ket{e}$ to $\ket{f}$ at a rate of $L_\text{Cl}$.
This leakage decays back partially into the computational subspace during each Clifford gate with a probability $1-d_q = 1-e^{-\tau_{\text{Cl}}\Gamma_{f\rightarrow e}}$.
The populations then evolve according to the rate equation
\begin{equation}
\vec{P}_\text{leak}(\text{n}_\text{Cl}+1) = M_{f\rightarrow e}  M_{\text{leak}} \vec{P}(\text{n}_\text{Cl}),
\end{equation}
where
\begin{equation}
M_{\text{leak}} = 
\begin{bmatrix}
    1-L_\text{Cl}/2 & L_\text{Cl} & 0\\
    L_\text{Cl}/2 &   1-L_\text{Cl}&0\\
    0&0&1
\end{bmatrix},
\end{equation}
captures the injected leakage and 
\begin{equation}
M_{f\rightarrow e} = 
\begin{bmatrix}
    1 & 1-d_q &0 \\
    0 &   d_q & 0\\
    0&0&1
\end{bmatrix}
\end{equation}
describes the decay back into the subspace.
The additional factor $1/2$ in the first column of $M_\text{leak}$ is due to the leakage injection only transferring population from $\ket{e}$ to $\ket{f}$ and the excited state being populated only half of the time, when averaging over all Clifford sequences.
Solving for the total leakage as a function of the number of Clifford gates $\text{n}_\text{Cl}$ yields
\begin{equation}
\label{eq:dynamics_A2_no_LR}
P_{f}^\text{leak}(\text{n}_\text{Cl}) = L_\text{Cl} \frac{1-\Big(e^{-\tau_{\text{Cl}}\Gamma_{f\rightarrow e}}\big(1-\frac{3}{2} L_\text{Cl}\big)\Big)^{\text{n}_\text{Cl}}}{2 e^{\tau_\text{Cl}\Gamma_{f\rightarrow e}} -2+3 L_\text{Cl}}.
\end{equation}
By taking the limit $\text{n}_\text{Cl}\rightarrow\infty$ we recover Eq.~\eqref{eq:A_2_no_LR} from the main text
\begin{equation}
P_{f}^\text{leak}(\infty)=A_2^\text{leak} = \frac{L_\text{Cl}}{3L_\text{Cl}+2\qty[e^{\tau_\text{Cl}\Gamma_{f\rightarrow e}}-1]}.
\end{equation}
The LR operation transfers most of the leakage to the resonator with fidelity $\mathcal{F}_\text{LR}$, where it leaves the resonator during one Clifford gate with probability $1-d_r = 1-e^{-(\tau_\text{Cl}+\tau_\text{leak}+\tau_\text{LR})\kappa_\text{R}}$.
\begin{figure}[h]
\includegraphics{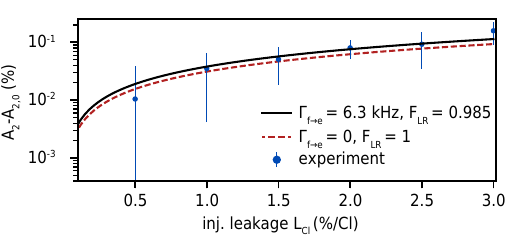} 
\caption{Rate equation prediction of the increase in RB steady-state leakage $A_2$ population with induced leakage per Clifford gate compared to experiment. The full rate equation taking into account the qubit decay from $\ket{f}\rightarrow\ket{e}$ and the LR success rate $\mathcal{F}_\text{LR}$ (solid black line) and the simplified model without second state decay and perfect LR operation (red dashed line) both fit to the measurement data (blue points) within error bounds.
\label{fig:rate_comp}}
\end{figure}

Therefore, the rate equation for the populations becomes
\begin{equation}
\vec{P}_\text{LR}(\text{n}_\text{Cl}+1) = M_\text{R} M_{f\rightarrow e} M_{\text{LR}} M_{\text{leak}} \vec{P}(\text{n}_\text{Cl}),
\end{equation}
with the added LR operation described by the matrix
\begin{equation}
M_{\text{LR}} = 
\begin{bmatrix}
    1 & \mathcal{F}_\text{LR} & -\mathcal{F}_\text{LR}/2\\
    0 &   1-\mathcal{F}_\text{LR}&\mathcal{F}_\text{LR}/2\\
    0&\mathcal{F}_\text{LR}& 1-\mathcal{F}_\text{LR}/2 \label{eq:A_LR}
\end{bmatrix},
\end{equation}
and the decay of the excitation in the resonator by the decay matrix
\begin{equation}
M_\text{R} = 
\begin{bmatrix}
    1 & 0 &0 \\
    0 &   1 & 0\\
    0&0&d_r
\end{bmatrix}.
\end{equation}
The last column in Eq.~\eqref{eq:A_LR} accounts for the effect of population from the resonator being swapped back into the higher qubit state $\ket{f}$.
For large $\text{n}_\text{Cl}$ the leakage population approaches
\begin{widetext}
\begin{equation} 
\label{eq:A_2_LR_full}
A_2^\text{LR} = \frac{{L_\text{Cl} e^{-\tau_\text{Cl}\Gamma_{f\rightarrow e}} (2 - 2F_\text{LR} + e^{-(\tau_\text{Cl}+\tau_\text{leak}+\tau_\text{LR})\kappa_\text{R}} (3 \mathcal{F}_\text{LR}-2))}}{{4 + 2 e^{-(\tau_\text{Cl}+\tau_\text{leak}+\tau_\text{LR})\kappa_\text{R}} (\mathcal{F}_{LR}-2) + (3 L_\text{Cl}-2) e^{-\tau_\text{Cl}\Gamma_{f\rightarrow e}} (2 - 2 \mathcal{F}_\text{LR} + e^{-(\tau_\text{Cl}+\tau_\text{leak}+\tau_\text{LR})\kappa_\text{R}} (3 \mathcal{F}_\text{LR}-2))}}.
\end{equation}
\end{widetext}
Assuming $\mathcal{F}_\text{LR}\approx 1$ and $\Gamma_{f\rightarrow e}\ll \kappa_\text{R}$ gives the limit imposed by the decay rate of the resonator stated in Eq.~\eqref{eq:A_2_LR_Assuming_TR_<Tef} of the main text,
\begin{equation} 
\label{eq:rate_Leak_LR_limit}
A_2^\text{LR} = \frac{L_\text{Cl} e^{-(\tau_\text{Cl}+\tau_\text{leak}+\tau_\text{LR})\kappa_\text{R}} }{4\qty[1- e^{-(\tau_\text{Cl}+\tau_\text{leak}+\tau_\text{LR})\kappa_\text{R}}]+3 L_\text{Cl}}.
\end{equation}
For the given device parameters, Eqs.~\eqref{eq:A_2_LR_full} and \eqref{eq:rate_Leak_LR_limit} differ only minimally, as shown in Fig.~\ref{fig:rate_comp}, suggesting that the effectiveness of the LR operation for repeated application is mostly limited by the decay rate of the resonator.

Next, we derive the dependency of the error rate when applying the LR operation after a leakage injection.
It is important to note that the LR operation does not remove leakage errors, it instead maps the leakage error to a Pauli error inside the computational subspace of the qubit.
The first-order error dependency of the error per Clifford gate on decoherent errors is~\cite{Abad2022}
\begin{equation}
\label{eq:error_dephasing}
    \varepsilon = \frac{\tau \Gamma_\Sigma}{3}.
\end{equation}
\begin{figure}[H]
\includegraphics{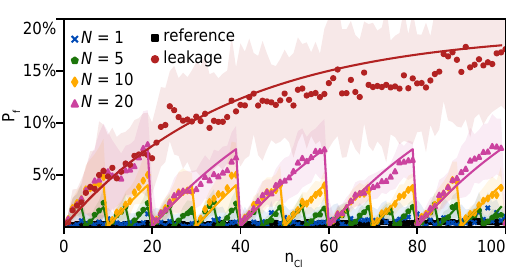} 
\caption{Population in $\ket{f}$ against the number of Clifford gates $\text{n}_\text{Cl}$, when the LR operation is applied every $N$ Clifford gates compared to a reference RB (black squares) and pure leakage RB (red circles). The shaded background is the standard deviation of the measurement data, the solid lines are predictions from rate equations taking into account the LR operation being applied only every 20, 10, 5, 1 Clifford gates (pink triangles, yellow diamonds, green pentagon, blue cross).
\label{fig:shark}
}
\end{figure}
\hspace{-4mm}Here, $\Gamma_\Sigma$ is sum of the relaxation rate $\Gamma_1$ and dephasing rate $\Gamma_\phi$ and $\tau$ the total time between Clifford gates, which includes in different cases the length of the Cliffords themselves $\tau_\text{Cl}$, the length of the leakage injection $\tau_\text{leak}$ and the length of the LR operation $\tau_\text{LR}$.
To get an expression for the effective dephasing rate due to leakage and subsequent recovery we begin with an arbitrary qubit-state in the subspace $\ket{\psi} = \alpha \ket{g}+\beta \ket{e}$. 
The density matrix including the first three levels for this state is given by
\begin{equation}
\rho=\left(\begin{array}{ccc}
|\alpha|^2 & \alpha \beta^* & 0\\
\alpha^* \beta& |\beta|^2 & 0\\
0 & 0& 0
\end{array}\right).
\end{equation}
The leakage injection with rate $L_\text{Cl}$ is described by the unitary matrix
\begin{equation}
R=\left(\begin{array}{ccc}
1 & 0 & 0\\
0& \sqrt{1-L_\text{Cl}} & \sqrt{L_\text{Cl}}\\
0 & \sqrt{L_\text{Cl}} & \sqrt{1-L_\text{Cl}}
\end{array}\right).
\end{equation}
This results in the density matrix $\rho'=R \rho R^\dagger$:
\footnotesize
\begin{equation}
\rho'= \left(\begin{array}{ccc}
|\alpha|^2 & \alpha \beta^* \sqrt{1-L_\text{Cl}} & \alpha \beta^* \sqrt{L_\text{Cl}}\\
\alpha^* \beta \sqrt{1-L_\text{Cl}}& |\beta|^2(1-L_\text{Cl}) & |\beta|^2 \sqrt{1-L_\text{Cl}}\sqrt{L_\text{Cl}}\\
\alpha^* \beta \sqrt{L_\text{Cl}} & |\beta|^2 \sqrt{1-L_\text{Cl}}\sqrt{L_\text{Cl}} & |\beta|^2 L_\text{Cl} 
\end{array}\right).
\end{equation}
\normalsize
After the application of an ideal LR operation all population in $\ket{f}$ is transferred to $\ket{e}$ without preserving the phase information due to the decoherent nature of the LR operation. 
This results in the final density matrix
\begin{equation}
\rho''=\left(\begin{array}{ccc}
|\alpha|^2 & \alpha \beta^* \sqrt{1-L_\text{Cl}} & 0\\
\alpha^* \beta \sqrt{1-L_\text{Cl}}& |\beta|^2 & 0\\
0 & 0& 0
\end{array}\right).
\end{equation}
We compare the off-diagonal elements of this matrix to the Bloch-Redfield model of decoherence~\cite{Wangsness1953, Bloch1957, Redfield1957} and get ${\sqrt{1-L_\text{Cl}} = e^{-\Gamma_\text{leak}\tau}}$. Assuming a small $L_\text{Cl}$ leads to an expression for the dephasing rate due to the decoherent recovery of leakage ${\Gamma_\text{leak} = L_\text{Cl}/(2 \tau)}$.
Inserting this expression in Eq.~\eqref{eq:error_dephasing} gives the error $\varepsilon \propto L_\text{Cl}/6$ as a function of leakage $L_\text{Cl}$ found in Eq.~\eqref{eq:error_LR} in the main text.

\begin{figure}[t!]
\includegraphics{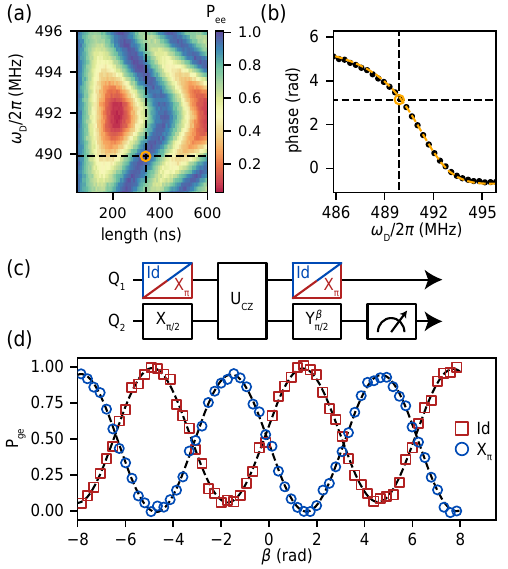} 
\caption{Calibration and characterization for the parametric controlled-Z gate. (a) Population of the $\ket{ee}$ state $P_{ee}$ when parametrically driving the transition $\ket{ee} \leftrightarrow \ket{fg}$ between qubit pair $\text{Q}_1$-$\text{Q}_2$ for varying drive frequency $\omega_\text{D}$ and length. The values used for the controlled-Z gate are indicated by the yellow marker. (b) Acquired phase difference between states $\ket{ee}$ and $\ket{ge}$ against the drive frequency $\omega_\text{D}$. (c) Pulse sequences for the measurement of the qubit phase. In the first sequence $\text{Q}_1$ stays in its ground state, in the second sequence $\text{Q}_1$ is excited. (d) Population $P_{ge}$ in the $\ket{ge}$-state against the rotation angle $\beta$ of the $Y_{\pi/2}$ pulse on the second qubit $\text{Q}_2$. In the two sequences $\text{Q}_1$ is prepared either in $\ket{g}$ (blue circles) or in $\ket{e}$ (red rectangles).
\label{fig:calib}
}
\end{figure}

\section{\texorpdfstring{\\* \vspace{2mm}}~Periodic application of LR}
\label{app:LR_ncl}
To compare the increase in leakage, when applying the LR operation every $N$ Clifford gates, we measure the population in $\ket{f}$ against the number of Clifford gates for $L_\text{Cl}=\SI{1}{\%}$ induced leakage for varying $N$. The result is shown in Fig.~\ref{fig:shark}.
When the LR operation is not applied (red circles) the leakage increases according to rate equation Eq.~\eqref{eq:dynamics_A2_no_LR}.
By performing the LR operation every $N$ Clifford gates the leakage population is periodically removed, producing a repeated shark fin pattern.
The solid lines following the data are predictions from the rate equation Eq.~\eqref{eq:dynamics_A2_no_LR}, where $\text{n}_\text{Cl}$ is replaced by $\text{n}_\text{Cl}(\text{mod }N)$.
The maximum leakage $ A_2^{\text{max}}$ with sparsely applied LR operations is upper bounded by the maximum slope of the leakage times the interval length $ A_2^{\text{max}} = N\cdot L_\text{Cl}/2$.
This opens up the possibility to apply fewer LR operations for devices with lower leakage, reducing the overall error from decoherence during device operation.

\section{\texorpdfstring{\\* \vspace{2mm}}~Controlled-Z gate calibration}
\label{app:CZ_calib}
The calibration of the controlled-Z gate follows the scheme presented in~\cite{Ganzhorn2020}.
We first excite both qubits and drive with a parametric pulse around the resonant frequency of the $\ket{ee} \leftrightarrow \ket{fg}$ transition, as shown in Fig.~\ref{fig:calib}(a).
From this measurement we determine the required drive length for one full oscillation for varying drive frequencies. 
Next, we measure the acquired phase difference between states $\ket{ee}$ and $\ket{ge}$ after one full oscillation [Fig.~\ref{fig:calib}(b)] using Ramsey sequences [Fig.~\ref{fig:calib}(c)].
From this measurement we extract the drive length $\tau_\text{CZ} = \SI{339}{\nano\second}$ and drive frequency $\omega_\text{D} = \SI{489.89}{MHz}$ for a $\pi$ phase, shown in Fig.~\ref{fig:calib}(d).
We calibrate and apply an additional virtual Z gate~\cite{McKay2017} on both qubits after the parametric pulse to account for the qubit frequency shift during the parametric drive.

\section{\texorpdfstring{\\* \vspace{2mm}}~ZZ Crosstalk Suppression}
\label{app:ZZ_crosstalk}
\begin{figure}[t!]
\includegraphics{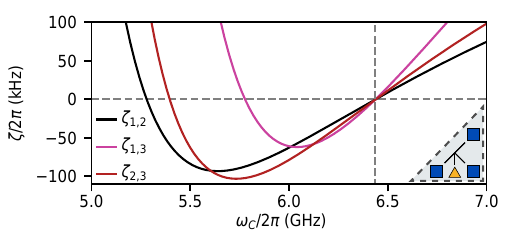} 
\caption{
ZZ-crosstalk between the three qubit pairs $\zeta_{12}$ (black), $\zeta_{13}$ (pink) and $\zeta_{23}$ (red) for varying coupler frequency using Eq.~\eqref{eqn:ZZ_fourth}. Full cancellation occurs at $\omega_C / 2\pi = \SI{6.44}{\giga\hertz}$, highlighted by the vertical dashed line.
\label{fig:crosstalk}
}
\end{figure}

Due to the anharmonicity of transmon qubits, their coupling results in a frequency shift of the $\ket{ee}$ state, causing unwanted ZZ interactions and degrading device performance.
For two qubits coupled via a tunable coupler, the strength of this ZZ crosstalk 
\begin{equation}
\zeta = E_{ee}-E_{eg}-E_{ge}
\end{equation}
is given by the shift in the energy level of the two-excitation state, $E_{ee}$, compared to the energies of the single-excitation states, $E_{eg}$ and $E_{ge}$~\cite{Sung2021}.
Using perturbation theory up to 4-th order~\cite{Li2020} its value is given by
\begin{equation}
\begin{aligned}
\zeta^{(4)}\left(\varphi_\text{ext}\right) &=\frac{2 g_{12}^2\left(\alpha_1+\alpha_2\right)}{\left(\Delta_{12}+\alpha_1\right)\left(\Delta_{12}-\alpha_2\right)}\\
& -\frac{2 g_{12} g_{1 c} g_{2 c}}{\Upsilon^2\left(\varphi_\text{ext}\right)}\left[\frac{1}{\Delta_2}\left(\frac{1}{\Delta_{12}}-\frac{2}{\Delta_{12}+\alpha_1}\right)\right. \\
& \left.+\frac{1}{\Delta_1}\left(\frac{2}{\Delta_{12}-\alpha_2}-\frac{1}{\Delta_{12}}\right)\right] \\
& -\frac{2 g_{1 c}^2 g_{2 c}^2}{\left(\Delta_1+\Delta_2-\alpha_c\right) \Upsilon^4\left(\varphi_\text{ext}\right)}\left(\frac{1}{\Delta_1}+\frac{1}{\Delta_2}\right)^2 \\
& +\frac{g_{1 c}^2 g_{2 c}^2}{\Delta_1^2 \Upsilon^4\left(\varphi_\text{ext}\right)}\left(\frac{2}{\Delta_{12}-\alpha_2}-\frac{1}{\Delta_{12}}+\frac{1}{\Delta_2}\right) \\
& +\frac{g_{1 c}^2 g_{2 c}^2}{\Delta_2^2 \Upsilon^4\left(\varphi_\text{ext}\right)}\left(\frac{2}{-\Delta_{12}-\alpha_1}+\frac{1}{\Delta_{12}}+\frac{1}{\Delta_1}\right),
\end{aligned}
\label{eqn:ZZ_fourth}
\end{equation}
where $\Delta_{12} = \omega_1-\omega_2$ denotes the detuning between the qubits, $g_{12}$ their coupling, $\Delta_i=\omega_C-\omega_i$ the detuning between qubits and coupler and $\Upsilon\left(\varphi_\text{ext}\right) = \left[E_{J}(0) / E_{J}\left(\varphi_\text{ext}\right)\right]^{1 / 4}$ a flux dependent rescaling to the couplings $g_{i c}$ between qubits and coupler. 
Crosstalk cancellation becomes possible when the qubit detuning is smaller than the qubit anharmonicity, a condition known as the straddling regime~\cite{Mundada2019}.
While this cancellation between a single qubit pair has been demonstrated~\cite{Sete2021floating, Sung2021, Li2020}, it is not immediately clear if crosstalk cancellation between more than two qubits connected to the same coupling element is possible.
This condition is fulfilled if the total crosstalk
\begin{equation}
\zeta_\Sigma =\sum_{i,j\in Q} |\zeta_{i,j}|,
\label{eq:app_zetasum}
\end{equation}
vanishes.
We show here, that this condition can be fulfilled in the parametric coupling architecture for the example of three qubits connected to the same coupling element.
We set the qubit frequencies to $\omega_1/2\pi = \SI{4.500}{\GHz}$, $\omega_2/2\pi = \SI{4.586}{\GHz}$, $\omega_3/2\pi = \SI{4.736}{\GHz}$, the anharmonicities to $\alpha_\text{Q}/2\pi = \SI{-240}{\MHz}$, $\alpha_\text{C}/2\pi = \SI{-160}{\MHz}$, the qubit-coupler interaction strength to be $g_{iC}/2\pi = \SI{110}{\MHz}$ and the qubit-qubit couplings to $g_{12}/2\pi = \SI{7.0}{\MHz}$, $g_{13}/2\pi = \SI{6.5}{\MHz}$, $g_{23}/2\pi = \SI{7.3}{\MHz}$.
For these values the individual contributions to $\zeta_\Sigma$ are suppressed at the same coupler frequency, as shown in Fig.~\ref{fig:crosstalk}.

We evaluate the influence of imprecise qubit frequency targeting by evaluating the minimum $\zeta_\Sigma$ against the coupler flux using Eq.~\eqref{eqn:ZZ_fourth}.
We use a Gaussian probability distribution of the qubit frequency with a standard deviation of $\SI{0.25}{\percent}$, a precision which can be achieved using laser annealing~\cite{Hertzberg2021}.
Randomly sampling over 10000 parameter sets, we extract an average minimum total crosstalk of $\overline{|\zeta_\Sigma|} = \SI{18}{\kHz}$.
Larger fluctuation of device parameters may be overcome by making one of the coupled qubits flux tunable.

\end{appendices}
\newpage
\bibliography{bib}

\end{document}